\documentclass[ljournal]{IEEEtran}
\usepackage{cite}
\usepackage{amsmath,amssymb,amsfonts}
\usepackage{algorithmic}
\usepackage{graphicx}
\usepackage{textcomp}
\def\BibTeX{{\rm B\kern-.05em{\sc i\kern-.025em b}\kern-.08em
    T\kern-.1667em\lower.7ex\hbox{E}\kern-.125emX}}

\usepackage{float}
\usepackage[hypertexnames=false]{hyperref}
\usepackage{cleveref}
\usepackage{multirow}
\usepackage{pgfplots}
\pgfplotsset{compat=1.7}
\usepackage{subcaption}
\usepackage{bm}
\renewcommand{\vec}[1]{\bm{#1}}
\def \r {{\bf r}}
\def \rr {{\bf r'}}
\def \rs {{\bf r_S}}
\def \rrs {{\bf r'_S}}

\def \nn {{\bf {n'}}}
\def \rns {{\bf {r_n^S}}}

\def \xnob {X_j^{\rm obs}}
\def \psisdata {\psi_{s}^{\rm data}}
\def \psidata {\psi_{\rm tot}^{\rm data}}
\def \psipred {\psi_{\rm tot}^{{\rm NN}(t)}}
\def \Nd{N_{\rm obs}}
\def \Ni{N_{\rm inv}}
\def \hpred{h^{{\rm NN}(t)}}
\def \psispred {\psi_{s}^{{\rm NN}(t)}}

\graphicspath{{./Figs/}}

\begin{document}
\title{Surface profile recovery from electromagnetic field with
physics--informed neural networks}

\author{Yuxuan Chen, Ce Wang, Yuan Hui, and Mark Spivack
\thanks{The paper was submitted on DD MM YYYY.}
\thanks{Y. Chen is with School of Mathematical Sciences, Soochow University, 
Suzhou, China (e-mail: chenyx@suda.edu.cn).}
\thanks{C. Wang is with Department of Electronic and Computer Engineering,
The Hong Kong University of Science and Technology, Hong Kong, China 
(e-mail: fogever@gmail.com).}
\thanks{Y. Hui is with Institute of Intelligent Computing Technology,
Chinese Academy of Sciences in Suzhou Branch, Suzhou, China
(e-mail: huiyuan.ict@gmail.com).}
\thanks{M. Spivack is with Department of Applied Mathematics and Theoretical 
Physics, University of Cambridge, Cambridge, UK (e-mail: ms100@cam.ac.uk).}
\thanks{Y. Chen, C. Wang and H. Yuan contributed equally to this work.}
\thanks{Corresponding author: Yuxuan Chen}
}

\maketitle


\begin{abstract}
Physics--informed neural networks (PINN) have shown their potential in
solving both direct and inverse problems of partial differential equations.
In this paper, we introduce a PINN-based deep learning approach to
reconstruct one-dimensional rough surfaces from field data illuminated by an
electromagnetic incident wave. In the proposed algorithm, the rough surface
is approximated by a neural network, with which the spatial derivatives of
surface function can be obtained via automatic differentiation and then the
scattered field can be calculated via the method of moments. The neural
network is trained by minimizing the loss between the calculated and the
observed field data. Furthermore, the proposed method is an unsupervised
approach, independent of any surface data, rather only the field data is
used.  Both TE field (Dirichlet boundary condition) and TM field (Neumann
boundary condition) are considered. Two types of field data are used here:
full scattered field data and phaseless total field data. The performance of
the method is verified by testing with Gaussian-correlated random rough
surfaces. Numerical results demonstrate that the PINN-based method can
recover rough surfaces with great accuracy and is robust with respect to a
wide range of problem regimes.
\end{abstract}


\section{Introduction}
\label{sec:introduction}

\IEEEPARstart{W}{ave} propagation from rough surfaces is of crucial importance
in the field of electromagnetic scattering and has been extensively studied.
The inverse problem of rough surface profile recovery from scattered
electromagnetic field remains a challenging problem with a wide range of
applications such as non--destructive testing, geophysics radar, remote
sensing, biomedical imaging and so many others. Much effort has therefore been
devoted to this field, for example~\cite{retov, chen2018rough1,
chen2018recovery, DESTOUCHES2001233, Akduman_2006, newton1, newton2, newton3,
newton4, sampling1, sampling2, sampling3, bao2016shape, dolcetti2021robust,
sefer1, sefer2, sefer3, iterative1, chen2018rough2, iterative3,
wombell1991reconstruction}. Common methods include the Rytov--type
methods~\cite{retov}, the Newton--type
methods~\cite{newton1,newton2,newton3,newton4}, reconstruction via sampling
(either linear sampling or other type)~\cite{sampling1,sampling2, sampling3},
and iterative methods with \emph{ad hoc}
guess~\cite{chen2018rough2,iterative1,iterative3}. These traditional methods
are generally based on physical equations of surface scattering, and their
robustness is contingent upon certain conditions on surface and incident
wave (e.g.~the convergence of iterative method in~\cite{chen2018rough2}
is based on low grazing angle approximation).

Recently, deep learning type methods have proven to be powerful approaches
dealing with a variety of problems, e.g.~image processing, natural
language processing, and scientific computing. Studies have been carried out
to recover physical scatters from the observed field data using a large
number of data,
see e.g.~\cite{dl1,dl2,dl3,dl4,dl5,dl6,dl7,dl8, dl9, cnn1,cnn2}.
In these works, neural networks are used to learn the mapping from the
observed electromagnetic field to the profile of target objects. In
particular, in~\cite{cnn1,cnn2}, a convolutional neural network (CNN) based
deep learning method has been developed to reconstruct rough surfaces from
the scattered data. It has been shown that deep learning methods are capable
of solving the inverse scattering problem in a more generalized way. On the
other hand, these deep learning methods are highly dependent on `solution'
data (ground truth). In the case of rough surface reconstruction, it is
difficult to obtain the entire profile of rough surfaces and corresponding
scattered data \emph{a priori}, especially when a large number of data are
required. Even if one can create data using simulation, the generated
geometric formulations may not be able to adequately capture the features of
real-world rough surfaces, resulting in a significant discrepancy in the
data's representation, which makes the trained model hard to employ in
practice.

Aiming to inherit the efficient learning strategy of data-driven methods
while maintaining the physics of rough surface scattering, it is worthwhile
to develop a unified method. This motivates us to employ the
physics--informed neural networks (PINN)~\cite{pinn1, pinn2}. PINN is a
particular type of deep learning approach developed to solve direct and
inverse problems of partial differential equations (PDEs). The main idea of
PINN is to employ neural networks as basis to approximate the `function'
(solution to PDE) and train the neural network by minimizing the `residual'
obtained from the physical information (PDEs), which can be realized using
the automatic differentiation. PINN is an \emph{unsupervised} learning
method, which does not require any \emph{a priori} knowledge of `solution'
data. PINN has now been widely applied to many practical engineering
problems; see for instance~\cite{pinn_direct1, pinn_direct2, pinn_direct3}
for solving the direct problem of electromagnetic scattering,
and~\cite{pinn_material1, pinn_is1, pinn_is2} for inverse scattering problems.

In this paper, we develop a novel deep learning method that is based on PINN
to reconstruct rough surfaces in two-dimensional media given the information
of field data. The physical information (equation) used in the method is the
integral representation relating the incident field, the scattered  field,
and the surface profile. Method of moments (MOM) is used to deal with
integral equations; it numerically calculates the scattered data from the
surface and incident field. A feed--forward neural network is served as basis
to approximate the surface height. The `predicted' scattered field is then
calculated by inserting the predicted surface (in terms of neural network)
and its spatial derivatives (obtained by automatic differentiation) to MOM.
The neural network is trained by minimizing the loss between the predicted
and observed field data. This renders an unsupervised training scheme
eliminating the need for rough surface data, instead solving the problem
using only the field data. The problem is solved using two types of observed
field data: full scattered data (including both strength and phase) and
phaseless total field data. In both cases, knowledge of incident field is
required. Both TE and TM fields are considered. We apply the method to
recover profiles of Gaussian-correlated random rough surfaces, and test the
method with a wide range of problem settings including various noise levels,
surfaces scales, surface heights, wavenumbers and angles of incidence. A
large number of numerical results validate the effectiveness and robustness
of the proposed approach.

The remainder of this paper is structured as follows. A brief review of
integral equations and method of moments is given in \cref{sec:background}.
In \cref{sec:recon}, we introduce a PINN-based deep learning approach to
reconstruct one-dimensional rough surfaces from field data, including the
detailed structure of neural networks and loss functions. Numerical
experiments are carried out in \cref{sec:examples}, in which the method is
tested with a range of problem regimes. Some concluding remarks and potential
further directions are drawn in \cref{sec:conclusions}.

\section{Background}
\label{sec:background}

Consider electromagnetic scattering from a perfectly electrically conducting
(PEC) 1D rough surface illuminated by an incident wave. The so--called
\emph{direct} problem is to calculate the scattered field given the rough
surface shape, whilst, the \emph{inverse} problem is to recover the profile
of rough surface from a set of measured scattered data. It is well known that
surface scattering can be formulated by integral equations, which have been
widely discussed~\cite{ie1,ie2}. The method of moments (MOM) is one of the
most common numerical techniques to numerically solve the integral
equations~\cite{mom1,mom2}. In this section, we give a brief review of
integral equations and method of moments.

\subsection{Integral Equations}
\label{subsec:ie}

Let the coordinate axes be $x$ and $z$, where $x$ is the horizontal and $z$ is
the vertical. We consider the time--harmonic scalar wave field $\psi(x,z)$,
which is time independent ($e^{-i\omega t}$ time dependence is discarded). The
governing equation for wave scattering is the Helmholtz equation
\begin{equation}
\nabla^2 \psi(x,z) + k^2 \psi(x,z)=0,
\end{equation}
where $k$ is the wavenumber, defined by $k=2\pi/\lambda$, in which $\lambda$
is the wavelength. Suppose that the 1D PEC rough surface, denoted by $z=h(x)$,
is at least second--order differentiable, i.e.~$h(x)\in C^2(\mathbb{R})$. It
has a compact support on $\mathbb{R}$, and perturbs from zero only inside the
domain $x\in[-L,L]$. Let $V$ be the upper homogeneous medium in which the
points lie above the surface, i.e.~$V=\{(x,z)|z > h(x)\}$, and $S$ be the set
of points lying on the surface with $S = \partial V = \{(x,z)|z=h(x)\}$.
We write $\r, \rr \in V$ as points in the medium, and $\rs, \rrs \in S$ for
the points along the surface. Denote the incident and scattered wave fields as
$\psi_i$ and $\psi_s$, respectively, with $\psi = \psi_i + \psi_s$ being the
total wave in the space.

The wave field in the upper free space $V$ can be formulated by the
Kirchoff--Helmholtz equation,
\begin{equation}
\begin{aligned}
&\psi (\r) = \psi_i(\r)\\
& + \int_{S} \nn \cdot \left[ \psi(\r) \nabla' G(\r; \rr)-G(\r; \rr)
\nabla' \psi(\rr) \right] ds',
\end{aligned}
\end{equation}
where $\nn$ is the unit normal vector to the surface, and $G(\r;\rr)$ is the
Green's function to the free space Helmholtz equation given by
\begin{equation}
G(\r; \rr)=\frac{i}{4}H_0^{(1)}\left(k|\r - \rr| \right),
\label{eqn:green}
\end{equation}
where $H_0^{(1)}$ is the zeroth--order Hankel function of the first kind.
The Kirchoff--Helmholtz equation can be transformed into a pair of coupled
integrals by applying boundary conditions. If the Dirichlet boundary
condition is assumed, namely,
$\psi = 0$ on $S$,
which corresponds to the transverse electric (TE) field impinging on a PEC
surface, then the coupled integrals become:
\begin{equation}
\psi_i(\rs) = \int_S G(\rs; \rrs) \frac{\partial \psi (\rrs)}{\partial \nn}ds',
\label{eqn:d1}
\end{equation}
and
\begin{equation}
\psi_s(\r) = -\int_S G(\r; \rrs) \frac{\partial \psi (\rrs)}{\partial \nn} ds',
\label{eqn:d2}
\end{equation}
where $\frac{\partial \psi} {\partial \nn}:= \nn \cdot \nabla' \psi$, noting
that the first equation is obtained via taking the limit as point $\r$
approaches the surface. On the other hand, if we assume the Neumann boundary
condition, namely,
$\vec{n} \cdot \nabla \psi = 0$ on $S$,
which corresponds to the transverse magnetic (TM) field applied on the
surface, then the coupled integrals become:
\begin{equation}
\psi_i(\rs)= \frac{1}{2}\psi (\rs) - \int_S \frac{\partial G(\rs; \rrs)}
{\partial \nn}\psi(\rrs)ds',
\label{eqn:n1}
\end{equation}
and
\begin{equation}
\psi_s(\r) = \int_S \frac{\partial G(\r; \rrs)}{\partial \nn}  \psi(\rrs) ds',
\label{eqn:n2}
\end{equation}
where $\frac{\partial G} {\partial \nn}:= \nn \cdot \nabla' G$, noting that
the term $1/2$ in the first equation results from the singularity of the
Green's function when limiting the point to the surface.

\subsection{Method of Moments}
\label{sec:mom}

Method of moments (MOM) is used to solve the direct problem, namely, obtain
the scattered data. It recovers the surface currents (surface derivative
$\partial \psi /\partial \vec{n}$ for TE field and total wave along surface
$\psi$ for TM field) from the first integral equation (\eqref{eqn:d1} or
\eqref{eqn:n1}) and calculate the scattered field from the second integral
(\eqref{eqn:d2} or \eqref{eqn:n2}). We first discretize the domain $[-L,L]$
uniformly into $N$ subintervals with $(N+1)$ points, and denote them as $x_l$
with $l=0,1,\ldots, N$, where $x_0=-L$ and $x_N=L$. The midpoint on each
subinverval $[x_{l-1}, x_{l}]$ is defined as $X_l = 1/2(x_{l-1}+x_l)$ for
$l = 1,2,\ldots, N$. In a typical physical situation, the scattered data in
the inverse problem are usually measured along a horizontal line in the
medium $V$. We choose to measure the scattered field along the line
$z=\zeta$, where $\zeta$ is a constant with $\zeta >h(x), \; x \in[-L,L]$.
Denote the discrete points along the surface as $\vec{r_n^S} = (X_n,h(X_n))$
and points along measurement line as $\vec{q_n} = (X_n,\zeta)$.


Consider first the Dirichlet condition (TE field). To the leading order, the
surface differential on each subinterval $[x_{l-1},x_l]$ can be treated by a
line segment, namely,
\begin{equation}
ds\approx dL = \sqrt{1+h'(x)^2}dx.
\label{eqn:sidff}
\end{equation}
We denote ${\bf \Psi_i}$ and ${\bf \Psi_s}$ as the vectors of surface incident
field and target scattered field in $\mathbb{C}^N$ with
$\vec{\Psi_i}[l] = \psi_i \left (X_l, h(X_l) \right)$
and
$\vec{\Psi_s}[l] = \psi_s(X_l, \zeta)$.
Then, the discretized form of integral equations for TE case (\eqref{eqn:d1}
and \eqref{eqn:d2}) gives rise to a linear system relating the surface incident
field and scattered field:
\begin{equation}
\vec{\Psi_s} = B_D A_D^{-1} \vec{\Psi_i},
\end{equation}
where $A_D \in \mathbb{C}^{N \times N}$ is given by
\begin{equation}
A_D[n,l] = \int_{x_{l-1}}^{x_l} G(\rns; \rrs) \sqrt{1+h'(x')^2}dx',
\label{eqn:ad}
\end{equation}
and $B_D \in \mathbb{C}^{N \times N}$ is given by
\begin{equation}
B_D[n,l] = \int_{x_{l-1}}^{x_l} G(\vec{q_n};\rrs) \sqrt{1+h'(x')^2}dx',
\label{eqn:bd}
\end{equation}
for $n,l=1,\ldots,N$. The terms in $A_D$ and $B_D$ are approximated by the
trapezium rule. By inverting the matrix $A_D$, the surface derivative can be
obtained, and using the calculated surface derivative, the scattered field can
be measured by a matrix--vector multiplication.


Similar treatment is applied to the Neumann boundary condition (TM field).
Again, there is a linear system relating the incident wave and scattered
wave from the discretized form of the integral equations for TM case
(\eqref{eqn:n1} and \eqref{eqn:n2}):
\begin{equation}
\vec{\Psi_s} = B_N A_N^{-1} \vec{\Psi_i},
\end{equation}
where $A_N \in \mathbb{C}^{N \times N}$ is given by
\begin{equation}
\begin{aligned}
&A_N[n,l] = \\
&\begin{cases}
\int_{x_{l-1}}^{x_l} \frac{\partial G(\rns; \rrs)}{\partial \nn} 
\sqrt{1+h'(x')^2}dx', \;\; n \neq l  \\
\frac{1}{2} - \int_{x_{n-1}}^{x_n} \frac{\partial G(\rns; \rrs)}
{\partial \nn}\sqrt{1+h'(x')^2}dx', \;\; n = l\\
\end{cases}
\label{eqn:an}
\end{aligned}
\end{equation}
and $B_N \in \mathbb{C}^{N \times N}$ is given by
\begin{equation}
B_N[n,l] = \int_{x_{l-1}}^{x_l}\frac{\partial G(\vec{q_n}; \rrs)}{\partial
\nn}\sqrt{1+h'(x')^2}dx',
\label{eqn:bn}
\end{equation}
for $n,l=1,\ldots,N$. The terms in $A_N$ and $B_N$ are calculated via the 
trapezium method.

In general, the scattered field along the line $z=\zeta$ can be written in 
the operator form of
\begin{equation}
\vec{\psi_s} = \text{MOM}[h(x), h'(x), h''(x)] := B_hA_h^{-1}\vec{\psi_i}^{h}.
\end{equation}
These operators are all dependent on the surface profile $h$ and its
derivatives.

\section{Reconstruction with physics--informed neural network}
\label{sec:recon}

\begin{figure*}[t!]
  \centering  \includegraphics[width=0.95\linewidth]{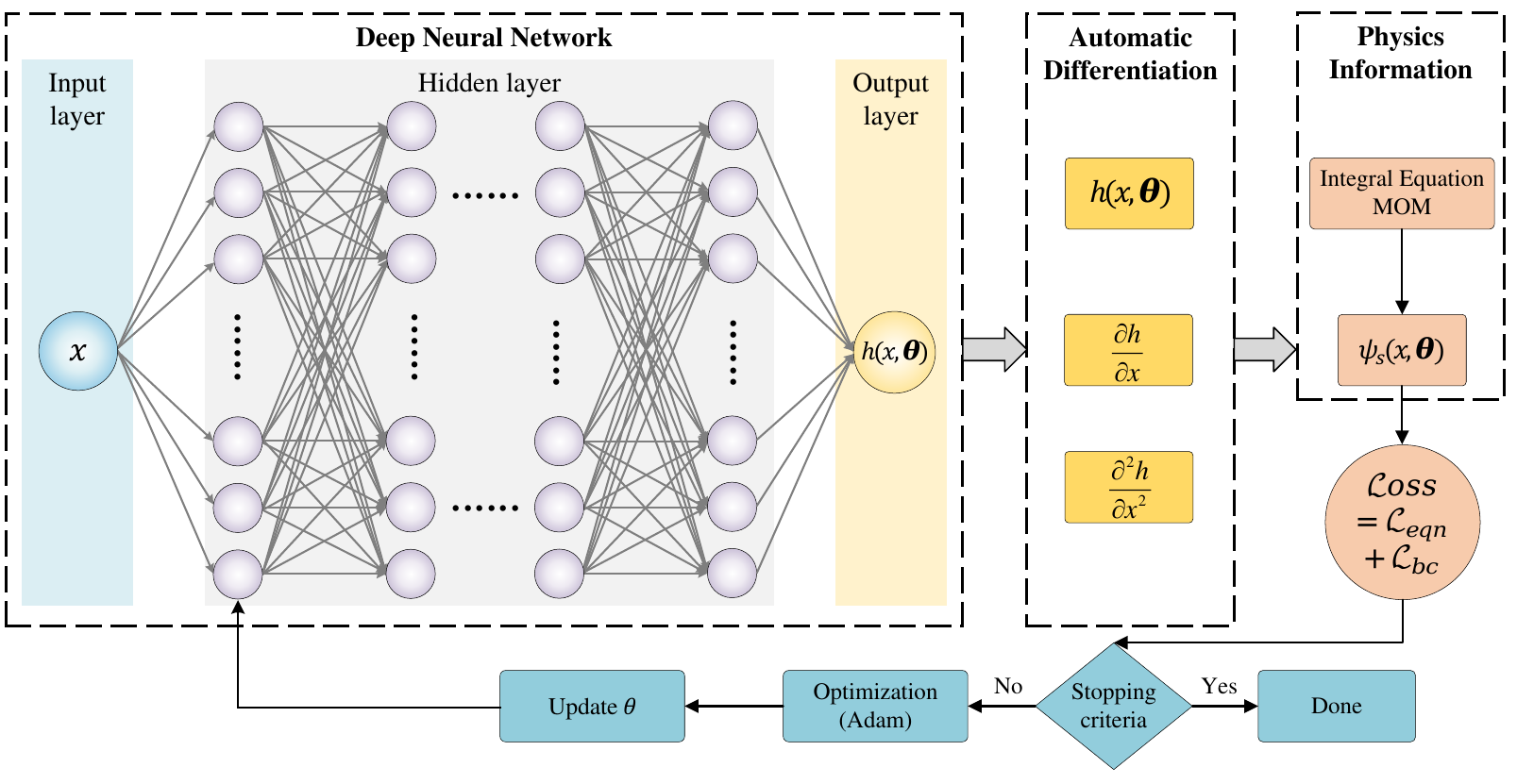}
  \caption{An illustrative figure for the PINN-based neural network to
  reconstruct rough surfaces from field data.}
  \label{fig:pinn_hei}
\end{figure*}

In this section, a novel unsupervised learning approach is proposed to
reconstruct one--dimensional rough surfaces from the field data, which is
based on physics--informed neural networks (PINN). The neural network is
utilized to approximate the surface height and trained based on the observed
field data. The architecture of the proposed learning method is presented in
\cref{fig:pinn_hei}. To illustrate the algorithm, we start with introducing
some preliminary knowledge (mesh, data, and boundary condition), followed by
the neural network structure, the physical information and loss function, and
finally the training scheme.
%
\subsection{Preliminary}
{\bf Sampling Points.} Suppose that we want to recover the rough surface
profile on the domain $[-L, L]$. At iteration (epoch) $t$ of training, we
first generate a random integer $N_t$ with $N_t\in[\Nd, \Ni]$ where $\Nd$ is
the size of observation data, and $\Ni$ is a larger number. The domain
$[-L,L]$ is discretized into $(N_t + 1)$ equally spaced points with $x_j^t =
j\Delta x_t-L$ for $j=0,1,\cdots,N_t$ and $\Delta x_t := 2L/N_t$. The
sampling points at $t$-th iteration of training are the midpoints $X_j^t$,
for $j=1,\ldots,N_t$ with $X_j^t = (x_{j-1}^t + x_j^t)/2$.

{\bf Field Data.}
Suppose that the field data is collected along a horizontal line $z=\zeta$
at observation points $(\xnob, \zeta)$ for $j=1,2,\cdots,\Nd$, where $\Nd$
is the number of observation points. Note that the observation points are not
aligned with the sampling points. Thus, a set of impulse (`hat--like')
piecewise linear functions are employed to interpolate the field data
from the observation points $(\xnob, \zeta)$ to the collocation points
$(X_j^t, \zeta)$.
Two types of field data are considered here. In the first case, we assume
that full field data can be detected, namely scattered field
$\psisdata(X_j^t, \zeta)$ is known. We also consider the case of phaseless
field data in which the amplitude of total field $\lvert \psidata
(X_j^t, \zeta) \rvert$ is observed with $\lvert \psidata \rvert= \lvert
\psisdata + \psi_i \rvert$.

{\bf Boundary Condition.}
The method also takes constraints on the boundary values. We consider that the
surface height at close-boundary points $h(X_j^b)$ for $j = 1,2,\ldots, N_b$
is known, where $X_j^b$ are close--boundary points (they are either close to
$L$ or $-L$). In the numerical examples of \cref{sec:examples}, a tapered
rough surface is used so that surface height close to edges decays to zero.

\subsection{Neural Network}
A fully connected feed--forward deep neural network is employed as basis to
approximate the surface function. The neural network is of uniform structure
which has a fixed number of hidden layers and neurons per hidden layer. A
nonlinear activation function is applied on the outputs after each hidden
layer. The input and output are set to be one-dimensional to predict the
function of the surface profile, with the input as one
spatial point and the output being the surface height at that point. At
iteration $t$ of training, we apply the neural network to all the sampling
points $X_n^t$ for $n=1,\ldots,N_t$, leading to the output of surface heights
on the entire domain. It is noted that we normalize the input data to the
range of $[-1,1]$, and the output, also conceptualized within the $[-1,1]$
range, is translated back to the target surface values through inverse
normalization.

\subsection{Physical Information and Loss Function}

At iteration $t$ of training, let $\hpred(x, \vec{\theta})$ be the predicted 
surface function in terms of neural network where $\vec{\theta}$ is the set of
parameters inside the neural network including the weights and biases. With
the predicted surface height in terms of neural network, the corresponding
spatial derivatives of surface,
$\partial \hpred(x,\vec{\theta}) / \partial x$ and
$\partial^2 \hpred(x, \vec{\theta}) / \partial x^2$,
can be obtained by automatic differentiation.
Then, the scattered field along the horizontal line $z = \zeta$ is calculated
from the predicted surface and its derivatives via the method of moments
(described in \cref{sec:mom}). This procedure can be viewed as combining the
neural network and physical equations of surface scattering. Denote the
calculated scattered field on the collocation points at $t$-th iteration of
training by $\psispred(X_n^t, \zeta, \vec{\theta})$, which takes the form of
\begin{equation}
\psispred =  \text{MOM}\left[\hpred, \frac{\partial\hpred}{\partial x},
\frac{\partial^2 \hpred}{\partial x^2} \right].
\end{equation}

If full scattered data is known, the loss function at iteration $t$ is given 
by a combination of two mean squared errors (MSE), MSE of scattered field
($\mathcal{L}_{\rm eqn}$) and MSE of boundary values ($\mathcal{L}_{\rm bc}$):
\begin{equation}
\begin{aligned}
&\mathcal{L}(\vec{\theta}, t) := \mathcal{L}_{\rm eqn} + \mathcal{L}_{\rm bc}
\\
&:= \frac{1}{N_t} \sum\limits_{j=1}^{N_t} \lvert \psispred(X_j^t, \zeta,
\vec{\theta}) - \psisdata(X_j^t, \zeta) \rvert^2 \\
&+ \frac{1}{N_b} \sum\limits_{j=1}^{N_b}\lvert \hpred(X_j^b, \vec{\theta}) -
h(X_j^b) \rvert^2.
\end{aligned}
\label{eqn:loss_full}
\end{equation}
Note that the first MSE of scattered field is complex--valued, it can be
thought of a combination of MSE of both real and imaginary parts, namely
\begin{equation}
\begin{aligned}
&\mathcal{L}_{\rm eqn} =\\
&\frac{1}{N_t} \sum\limits_{j=1}^{N_t} \lvert  \Re\{\psispred(X_j^t, \zeta,
\vec{\theta})\}- \Re\{\psisdata(X_j^t, \zeta)\}  \rvert^2\\
&+\lvert  \Im\{\psispred(X_j^t, \zeta, \vec{\theta})\}- \Im\{\psisdata(X_j^t,
\zeta)\}  \rvert^2.
\end{aligned}
\end{equation}

In the case of known phaseless total field data, the loss function at
iteration $t$ is composed of MSE of total field data
($\mathcal{L}_{\rm eqn}$) and MSE of boundary values ($\mathcal{L}_{\rm bc}$):
\begin{equation}
\begin{aligned}
&\mathcal{L}(\vec{\theta}, t) := \mathcal{L}_{\rm eqn} + \mathcal{L}_{\rm bc}\\
& := \frac{1}{N_t} \sum\limits_{j=1}^{N_t}\left \lvert |\psipred(X_j^t, \zeta, \vec{\theta})|  - |\psidata(X_j^t, \zeta)| \right \rvert^2\\
& + \frac{1}{N_b} \sum\limits_{j=1}^{N_b} \lvert \hpred(X_j^b, \vec{\theta}) - h(X_j^b) \rvert^2,
\end{aligned}
\label{eqn:loss_phaseless}
\end{equation}
where $\psipred = \psispred + \psi_i$ is the total field obtained by MOM
using predicted surface at iteration $t$ of training.

\subsection{Training}
Finally, the loss function is minimized by an optimizer to achieve an
`optimal' set of parameters $\vec{\theta}$. In this paper, the optimizer is
taken to be the Adam (Adaptive Moment Estimation) \cite{kingma2014adam}.
During the optimization, a learning rate is used to control how much the
weights of neural network are adjusted with respect to the gradient of the
loss function. After training, the neural network serves as an approximate
function to the rough surface, where inputting point $x$ yields the output
$h(x)$, the predicted surface height at that point.

\section{Numerical examples}
\label{sec:examples}

The performance of the proposed  method is tested here. In all numerical
examples, the incident field is taken to be a plane wave with $\psi_i(x,z) =
\exp \left(ik (\cos \alpha x + \sin \alpha z) \right)$, where $k$ is the
wavenumber and $\alpha$ is the grazing angle (complementary to the angle of
incidence). The domain of rough surface is set as $[-L, L]$ with $L=8\lambda$
where $\lambda$ is the wavelength of incident wave. We apply the method to
random Gaussian-correlated rough surfaces, which are generated by an
autocorrelation function (a.c.f) given by
\begin{equation}
\rho(\eta) = \langle h(x)h(x+\eta) \rangle = \exp\left(-\frac{\eta^2}
{l^2}\right),
\label{eqn:acf}
\end{equation}
where $l$ is the surface scale (autocorrelation length). The surface is
tapered to zero as $x$ approaches the edges ($x=L$ and $x=-L$) via a tanh
function. All the rough surfaces have mean zero and we scale the generated
surface vertically to vary the peak-to-trough height ($h_{\max}-h_{\min})$.
The field data is measured along the horizontal line $z=\zeta$, and the
number of observation points is set to be $\Nd$. We consider two cases here:
\begin{trivlist}
\item {\bf Case A}: full scattered data is known.
\item {\bf Case B}: phaseless total field data (amplitude of field) is known.
\end{trivlist}
If the measured field at one point is $\psi_{\rm data}$, then the noisy data
is given by
\begin{equation}
\psi_{\rm noise} = \psi_{\rm data} (1 + \epsilon\vartheta),
\end{equation}
where $\epsilon$ is the noise level and $\vartheta$ is a random number
uniformly generated in $[-1,1]$. Both TE field (Dirichlet boundary condition) 
and TM field (Neumann boundary condition) are considered. We test the method
with respect to several characteristics of rough surface and incident field
including surface scale, peak-to-trough height, wavenumber, and angle of
grazing. Throughout
\cref{sec:example_withoutnoise,sec:example_withnoise,sec:example_scale,sec:example_height},
the incident field is taken to be the plane wave with wavenumber $k=2\pi$
(corresponding to frequency of $300$MHz and wavelength of $1$ meter) and
grazing angle $\alpha = -\pi/4$.

As discussed in \cref{sec:recon}, uniform neural networks are employed with
$N_l$ hidden layers and $N_n$ neurons per hidden layer and an activation
function applied after each hidden layer. The structures of neural networks
as well as the optimization parameters are given in \cref{table:structure}
for TE and TM fields and case A and case B.
\begin{table}[b]
\centering
\caption{Structures of neural networks and optimization parameters}
\label{table:structure}
\begin{tabular}{c|cc|cc}
\hline
\multirow{2}{*}{Field} & \multicolumn{2}{c|}{TE}     
& \multicolumn{2}{c}{TM}      \\ \cline{2-5} 
  & case A     & case B    & case A     & case B    \\
\hline
$N_l$     & 4          & 6         & 4          & 8         \\
$N_n$     & 256        & 512       & 256        & 512       \\
activation function    & \multicolumn{2}{c|}{Sigmoid} &  \multicolumn{2}{c}{Sigmoid}     \\
\hline
optimizer         & \multicolumn{2}{c|}{Adam} &  \multicolumn{2}{c}{Adam}   \\
learning rate     & \multicolumn{2}{c|}{0.001} & \multicolumn{2}{c}{0.001}    \\
number of iterations  & 1500       & 2000      & 1500       & 2000    \\
\hline
\end{tabular}
\end{table}
The number of sampling points at each iteration is randomly selected from
$[\Nd, \Ni]$, where $\Ni$ is a larger number. For the boundary terms in the
loss function ($\mathcal{L}_{\rm bc}$ in \eqref{eqn:loss_full} and
\eqref{eqn:loss_phaseless}), we take $N_b=10$, and suppose that $h(X_j^b)=0$
with $x_j^b = -L+j\Delta x_t$ for $j = 0,1,\ldots,4$, and
$X_j^b = L - (j-5)\Delta x_t$ for $j = 5, 6, \ldots, 9$.

The quantitative error tested here is the $\ell^2$-norm error between
recovered and actual surfaces, given by
\begin{equation}
\lVert e \rVert_{\ell^2} = \frac{\lVert \vec{H^{\rm NN}} -\vec{H}
\rVert_{\ell^2}}{\lVert\vec{H} \rVert_{\ell^2}} \times 100\%
\end{equation}
where $\vec{H^{\rm NN}}$ is the vector of reconstructed surface and $\vec{H}$
is the vector of original surface. To analyze the error, we conduct $50$
tests and record both the mean and the standard deviation of the $\ell^2$-norm
error values.

All the examples are computed using Python. The functionality of neural
networks is supported by the open--source library PyTorch~\cite{pytorch}. The 
automatic differentiation is also implemented using PyTorch~\cite{automatic}. 
The source code of implementation is freely available in the supporting 
material \cite{supporting-material}.

\subsection{Reconstruction with noiseless data}
\label{sec:example_withoutnoise}

\begin{figure}[t]
  \centering
  \begin{subfigure}{0.5\columnwidth}
    \centering
    \includegraphics[width=0.9\linewidth]{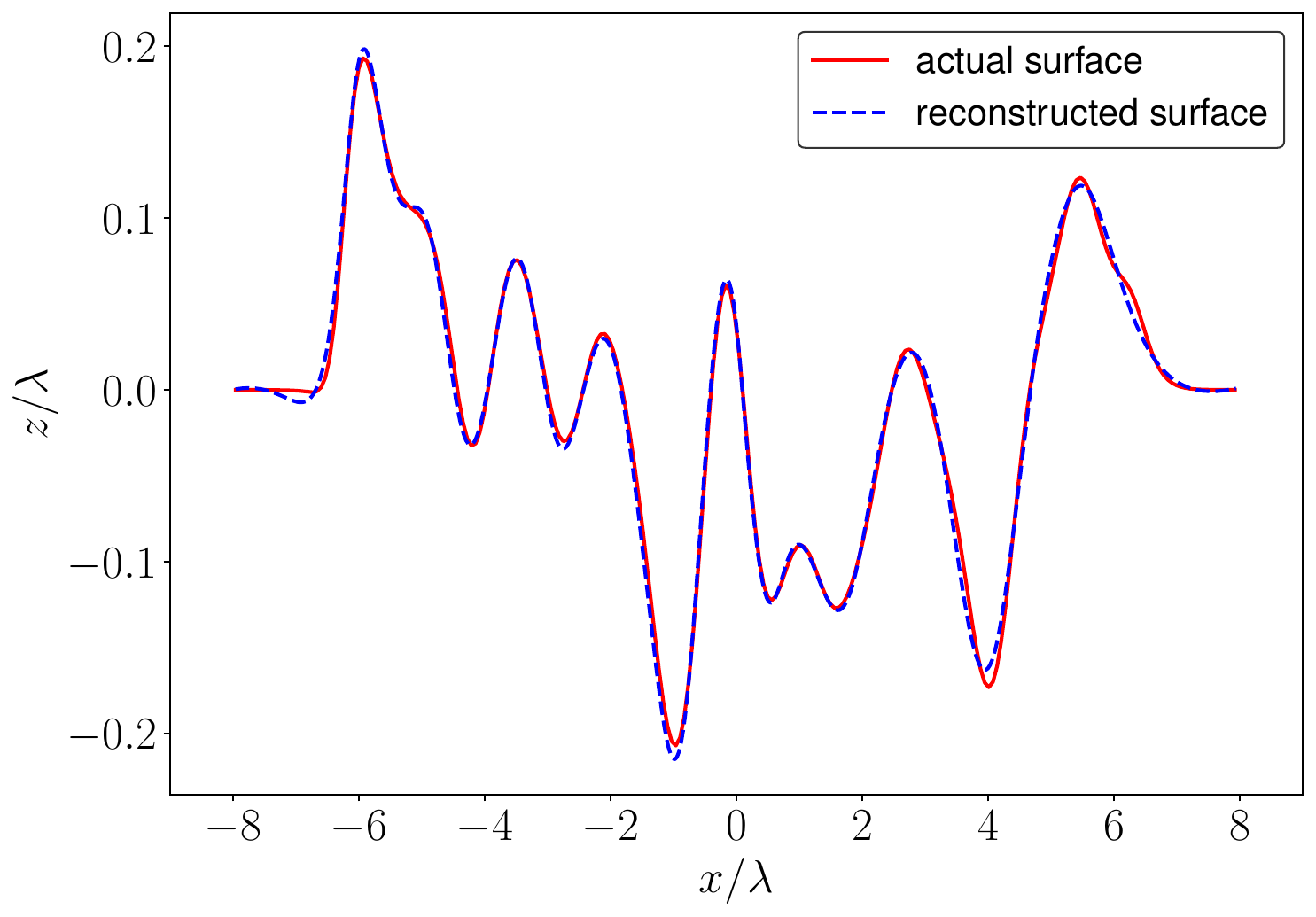}
    \subcaption{TE field}
  \end{subfigure}%
  \begin{subfigure}{0.5\columnwidth}
    \centering
    \includegraphics[width=0.9\linewidth]{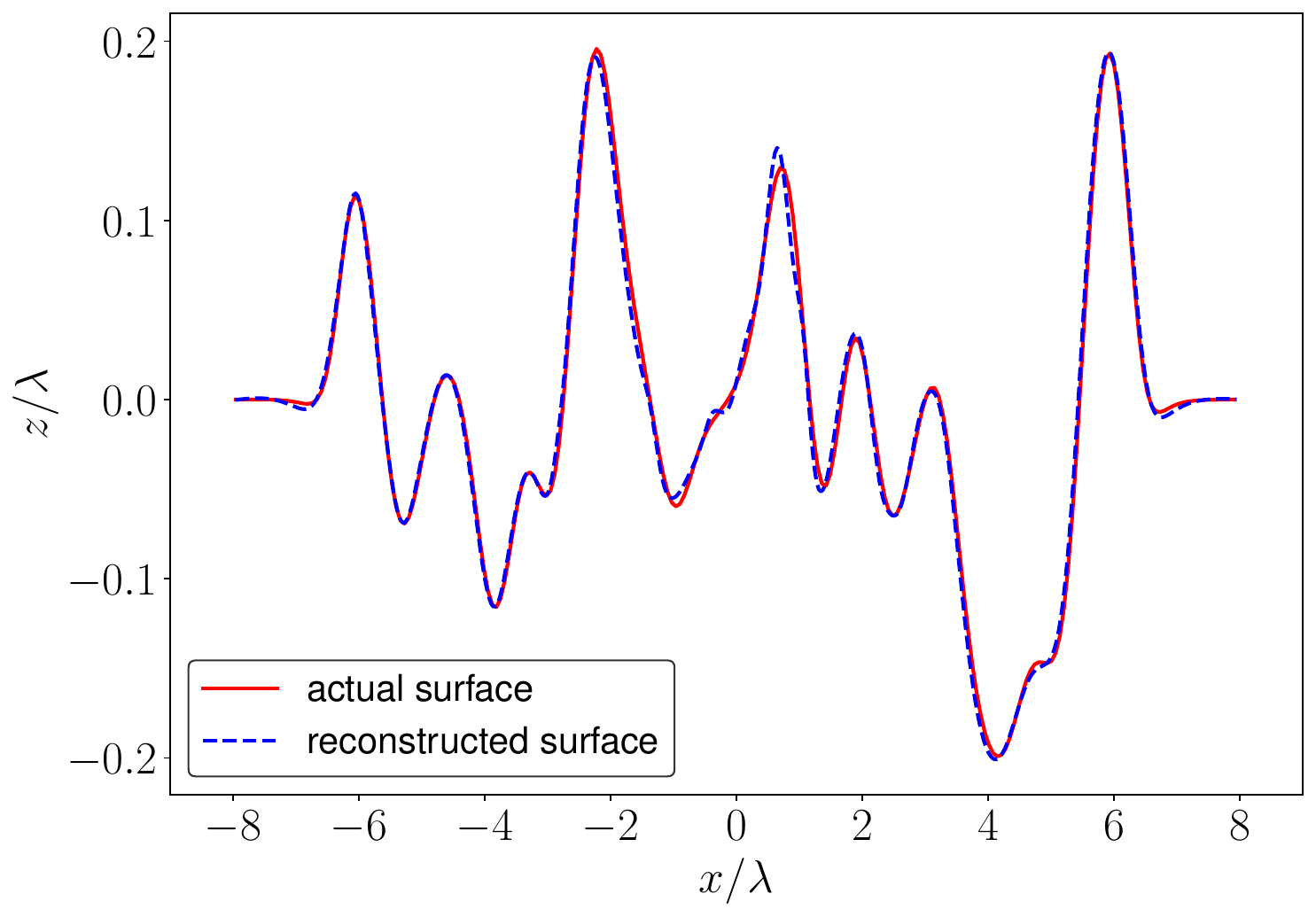}
    \subcaption{TM field}
  \end{subfigure}
  \caption{Reconstruction of rough surfaces against the actual surfaces for 
  case A (known full scattered data) without noise.}
  \label{fig:ex1_fulldata_recons}
\end{figure}

We first test the method using the field data without noise. The surface scale
is taken to be $l=2/3\lambda$, peak-to-trough height is $0.4 \lambda$, and the
data observation height is $\zeta = 0.5 \lambda$. Here we take $\Nd = 240$,
approximately $10$ points per surface scale, and $\Ni = 480$, approximately
$20$ points per surface scale. \Cref{fig:ex1_fulldata_recons} shows the
recovered surface plotted against the true surface for case A (known full
scattered data) without noise for TE and TM fields.
\begin{figure}[t]
  \centering
  \begin{subfigure}{0.5\columnwidth}
    \centering
    \includegraphics[width=0.9\linewidth]{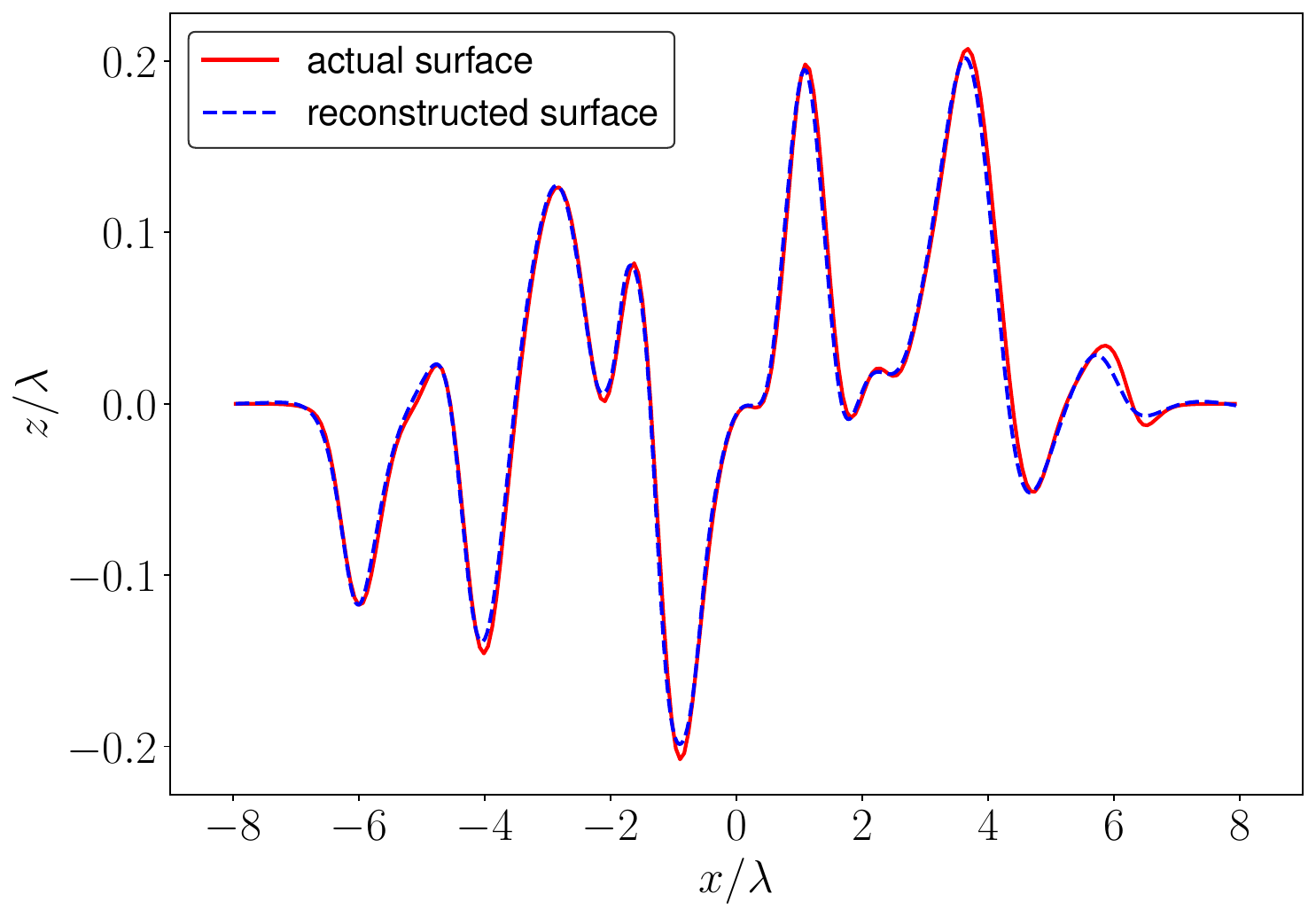}
    \subcaption{TE field}
  \end{subfigure}%
  \begin{subfigure}{0.5\columnwidth}
    \centering
    \includegraphics[width=0.9\linewidth]{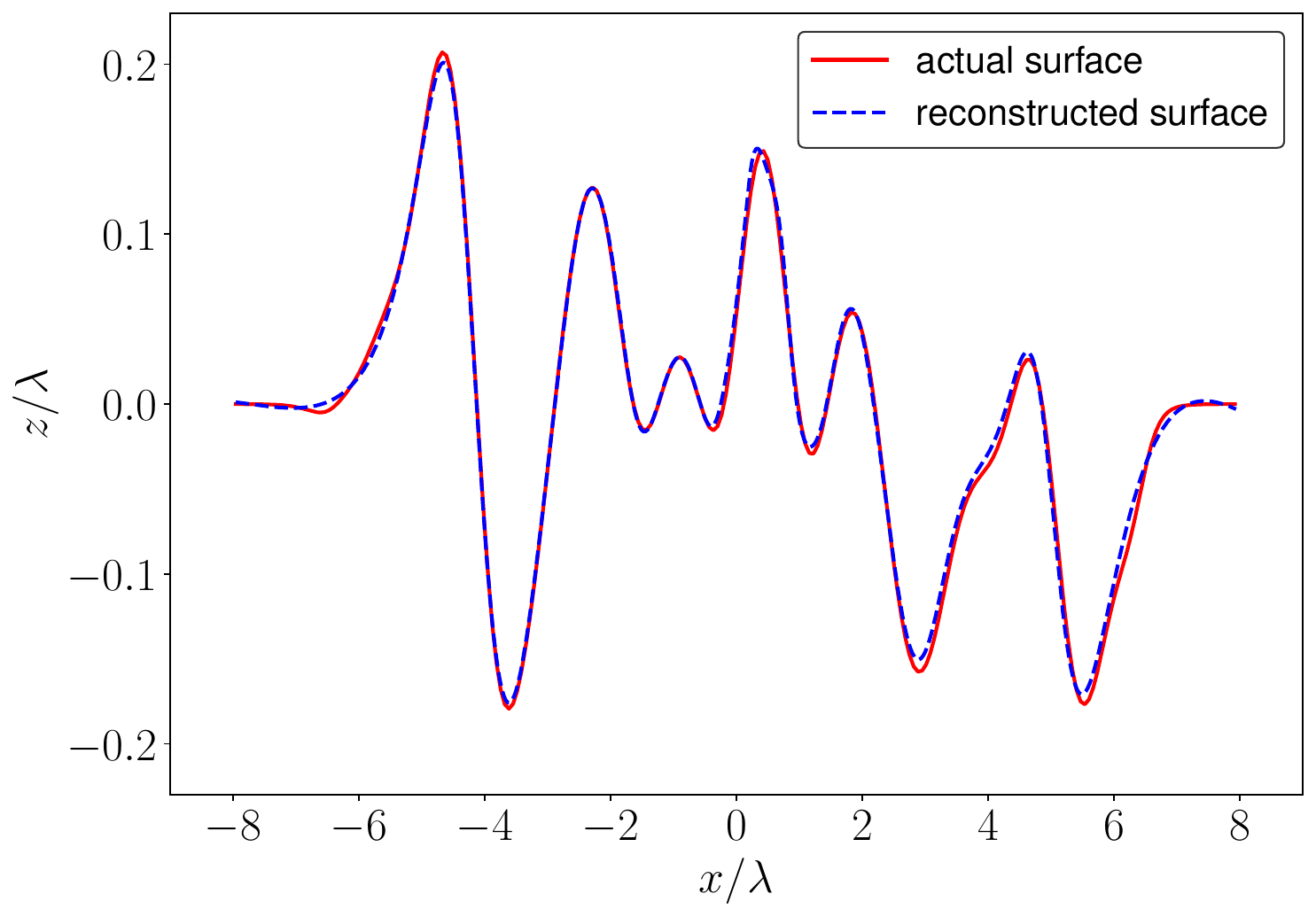}
    \subcaption{TM field}
  \end{subfigure}
  \caption{Reconstruction of rough surfaces against the actual surfaces
  for case B (known phaseless total field data) without noise.}
  \label{fig:ex1_phasedata_recons}
\end{figure}
The recovered surface matches well with the actual surface by capturing all
the shapes. In most parts of the surface, the recovered and actual surfaces
closely coincide. The main discrepancy occurs at peaks and troughs of the
surface. This is due to the significant change in the derivative of surface,
and has also been observed in the results obtained by other traditional 
methods~\cite{chen2018rough1,chen2018rough2}.

\begin{figure*}[t!]
  \centering
  \includegraphics[width=0.8\textwidth]{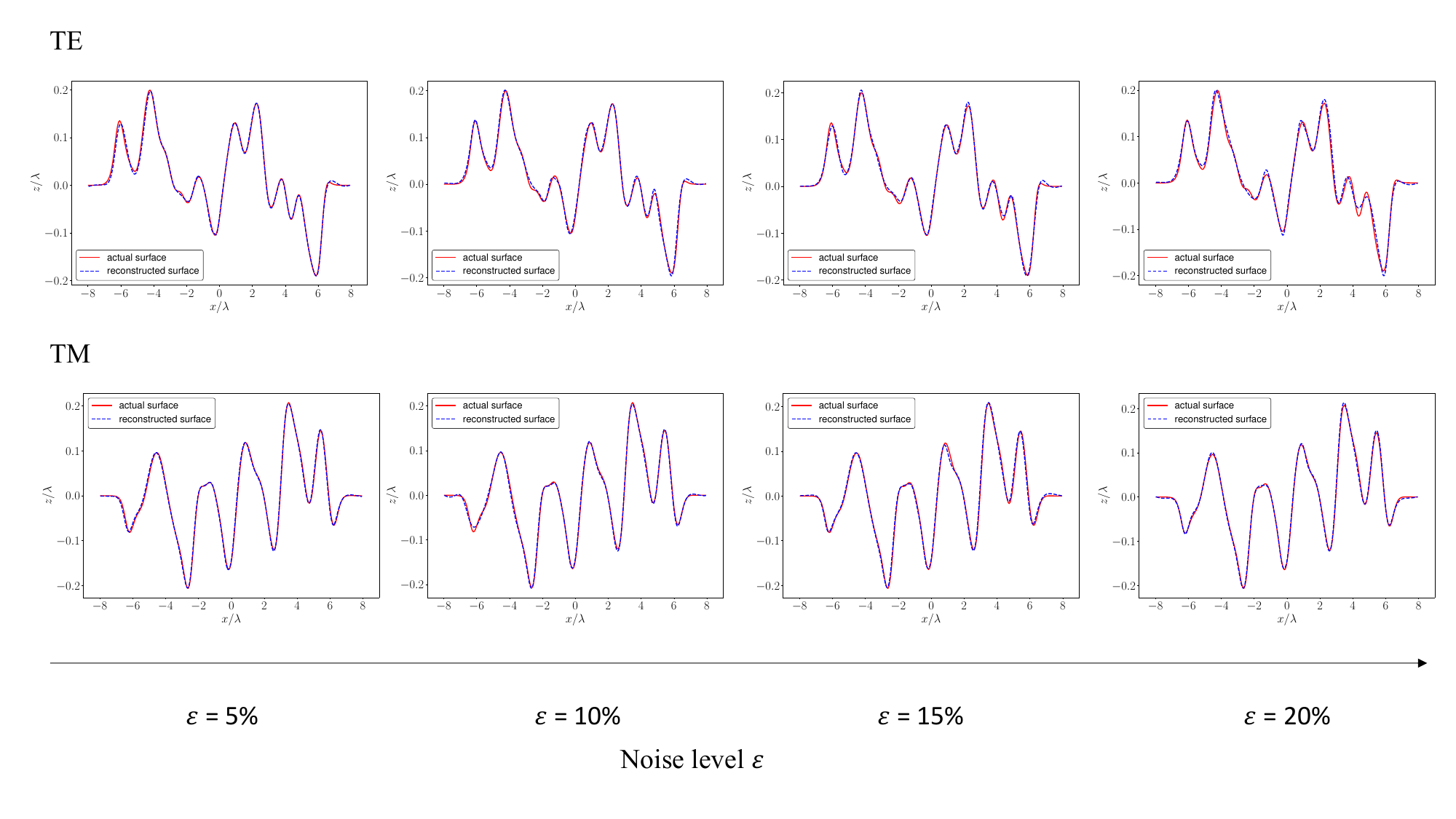}
  \caption{Reconstruction of rough surfaces compared to actual surfaces
using data with different noise levels $\epsilon$ in the case A of known 
scattered field data for TE and TM fields.}
  \label{fig:ex2_fullnoise_recons}
\end{figure*}
\begin{figure*}[t!]
  \centering
  \includegraphics[width=0.8\textwidth]{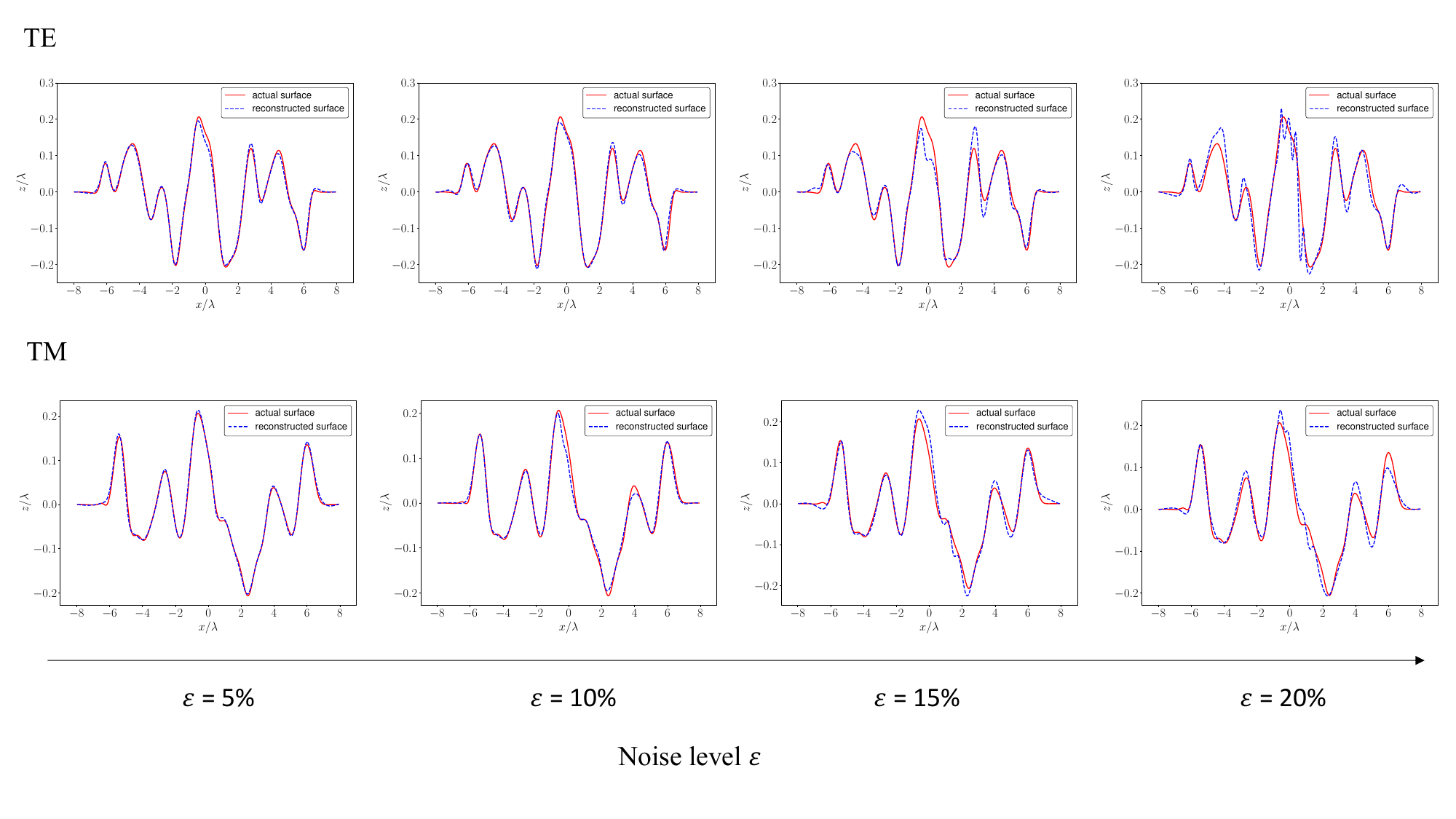}
  \caption{Reconstruction of rough surfaces compared to actual surfaces with
  respect to different noise levels $\epsilon$ in the case B of known 
  phaseless total field data for TE and TM fields.}
  \label{fig:ex2_phasenoise_recons}
\end{figure*}

For case B where the amplitude of total field is known, the resulting
solution versus original surface with non-noisy data for TE and TM fields is
shown in \cref{fig:ex1_phasedata_recons}. Similar to the previous case, the
solution surface closely matches the actual surface.

\subsection{Reconstruction with noisy data}
\label{sec:example_withnoise}

Here we add uniformly distributed white noise to the field data, and test the
results with respect to noise level $\epsilon$. The problem setting is kept
the same as in \cref{sec:example_withoutnoise} ($l=2/3\lambda$,
$h_{\max}- h_{\min} = 0.4 \lambda$, $\zeta = 0.5 \lambda$, $\Nd = 240$,
and $\Ni = 480$).  The recovered surfaces against the original surfaces with
respect to a range of noise levels ($\epsilon$) are shown in
\cref{fig:ex2_fullnoise_recons} for TE and TM fields in case A (full
scattered field known).
Clearly, with noisy data, the recovered surface still has close agreement with
the actual surface, which validates the robustness of the method in the
presence of noise. The discrepancies become more apparent when using data
with larger noise level. Surprisingly, for case A where full scattered data
is known, this method can tolerate noise with noise level up to $20\%$.

Now we consider the performance in case B (phaseless total field known).
The recovered surface plotted with the original surface with respect to
noise level $\epsilon$ is shown in \cref{fig:ex2_phasenoise_recons} for TE and
TM fields.
It is found that the reconstruction using phaseless data with small noise 
level still matches well with the actual surface. As the noise level 
increases, in particular when $\epsilon > 5\%$, clear discrepancies show up, 
leading to much larger error. However, all the `solution' surfaces recapture 
the overall shape of original ones.

\begin{figure}
  \centering
  \begin{subfigure}{0.9\linewidth}
    \centering
    \includegraphics[width=0.75\linewidth]{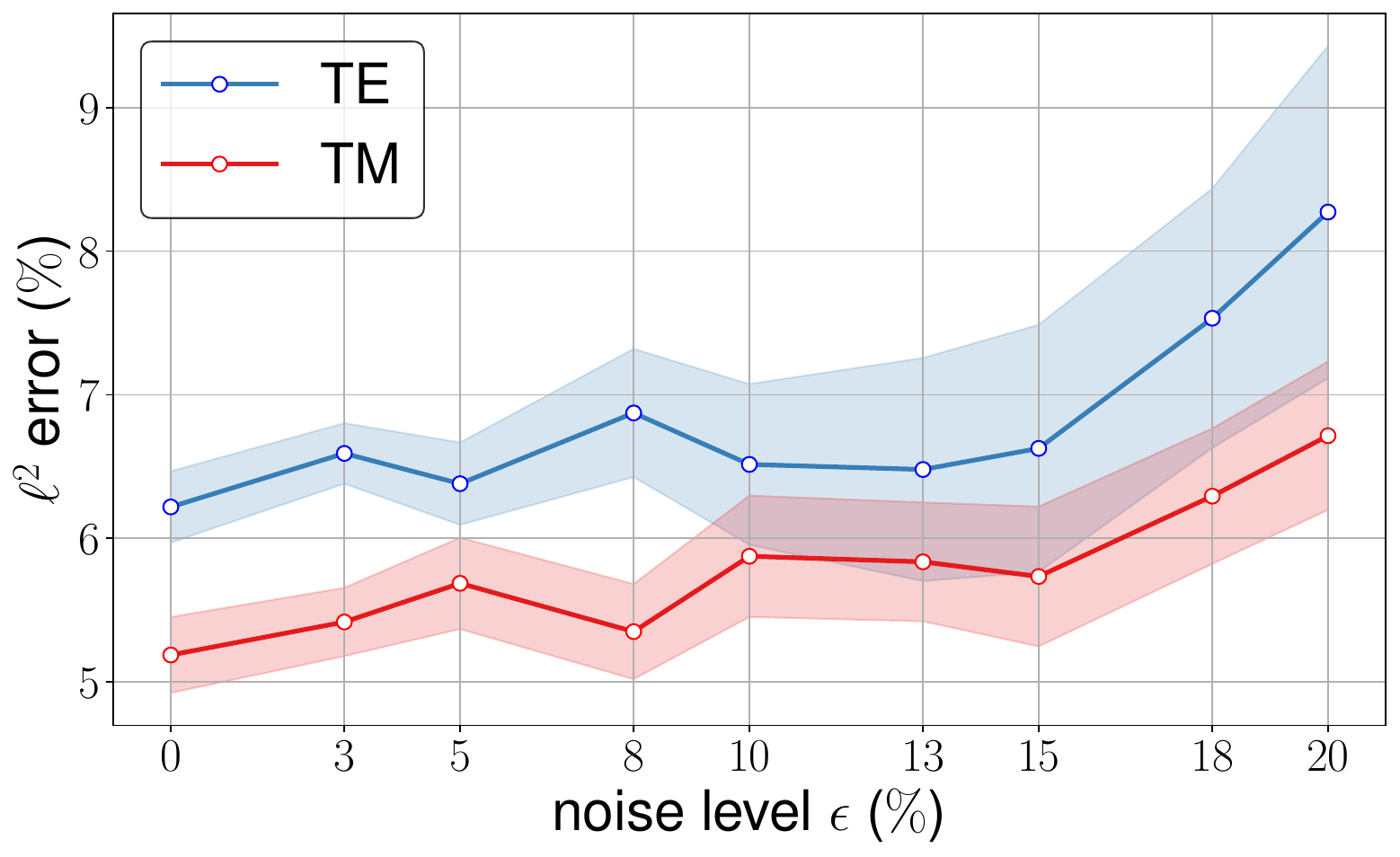}
    \subcaption{case A}
  \end{subfigure}%
  \vspace{0.2cm}
  \begin{subfigure}{0.9\linewidth}
    \centering
    \includegraphics[width=0.75\linewidth]{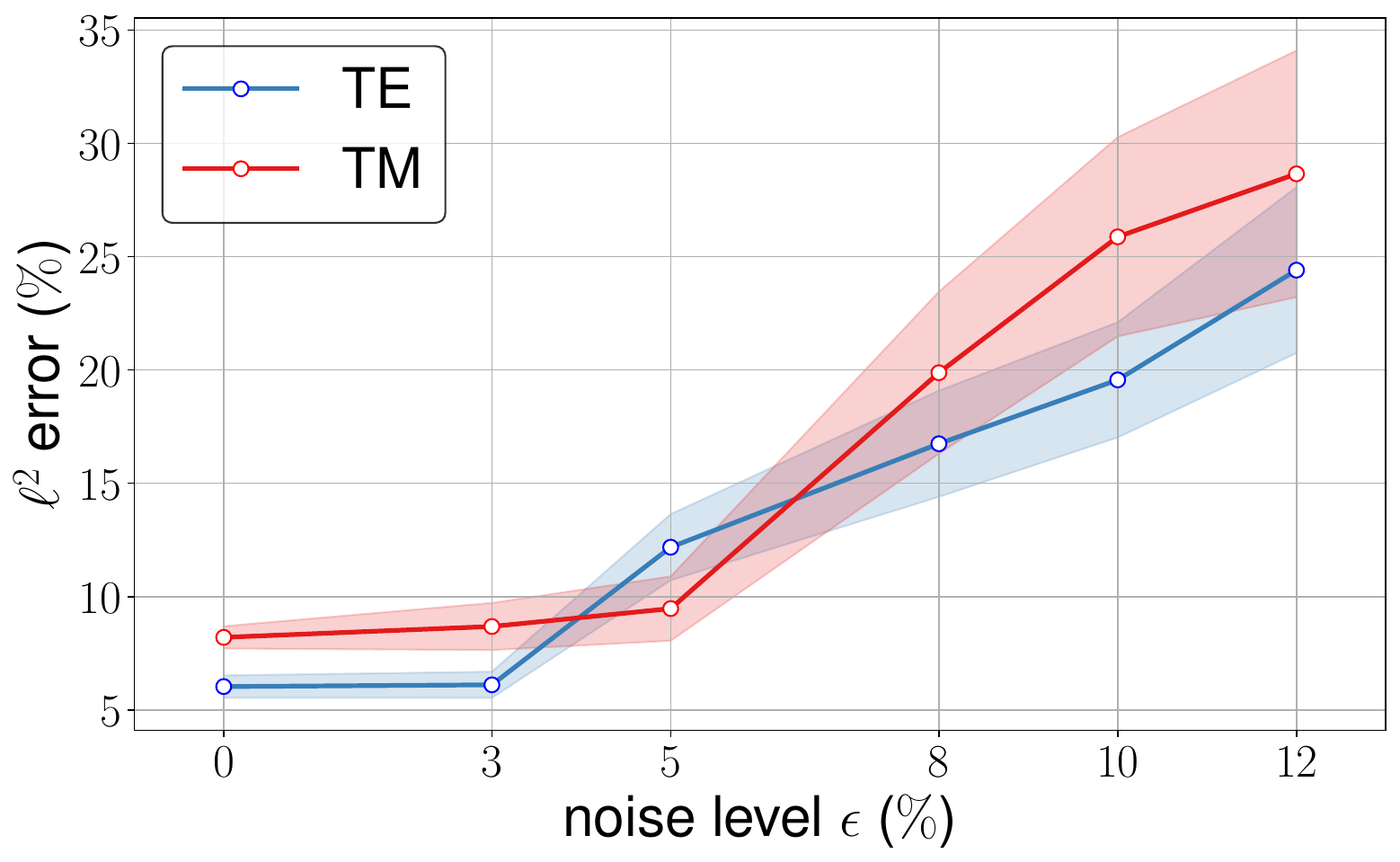}
    \subcaption{case B}
  \end{subfigure}
  \caption{Mean $\ell^2$ norm error (points) and standard deviation (colored 
  band) with respect to noise level ($\epsilon$) for case A (full scattered 
  data) and case B (phaseless total field data).}
  \label{fig:ex2_error}
\end{figure}

We also analyze the error quantitatively. The mean $\ell^2$-norm error
together with the standard deviation with respect to noise level ($\epsilon$) 
is presented is \cref{fig:ex2_error} for case A and case B.
For the case A with full scattered data, the error stays at a similar level 
when noise is small, whereas it rises significantly when noise level exceeds 
$15\%$. This phenomenon is more pronounced for TE field. In the case A, the
standard deviation also keeps small for small noise level, but it increases
significantly when noise level is large. In the case B with phaseless data,
there is a steady increment in the error with noise level for both TE and TM
fields. In summary, the method is able to tolerate noise level $\epsilon$ up 
to around $15\%$ for case A and $5\%$ for case B.

Throughout the remainder of the numerical examples, we will add $10\%$ noise 
to the data in case A (full scattered data) and $3\%$ noise to the data in
case B (phaseless total field).

\subsection{Reconstruction with respect to surface scale}
\label{sec:example_scale}

\begin{figure}
\captionsetup[subfigure]{justification=centering}
\centering
\begin{subfigure}{0.5\columnwidth}
\centering
\includegraphics[width=0.9\linewidth]{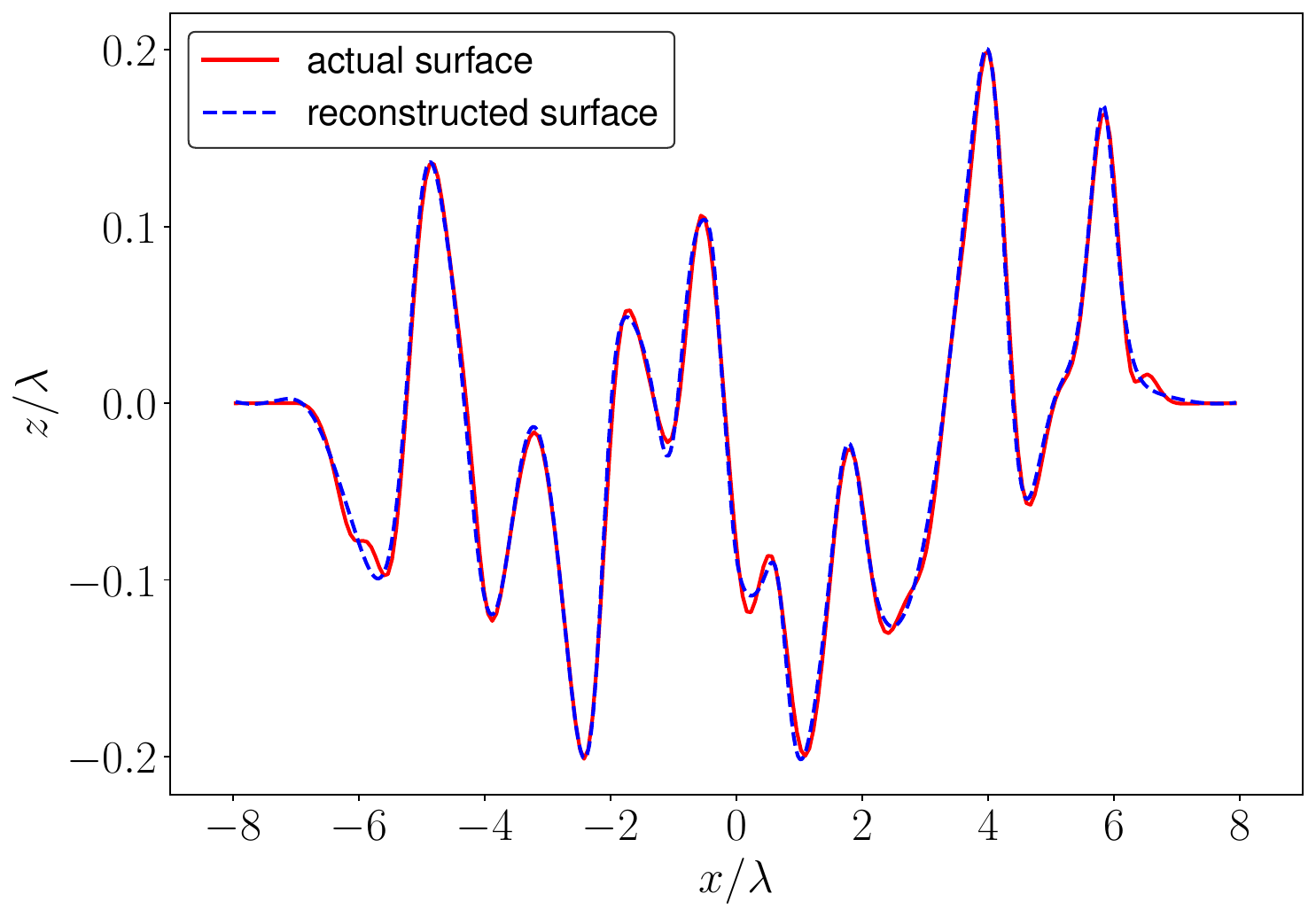}
\subcaption{case A, TE, $l=\frac{1}{2}\lambda$, $\Ni = 600$}
\end{subfigure}%
\begin{subfigure}{0.5\columnwidth}
\centering
\includegraphics[width=0.9\linewidth]{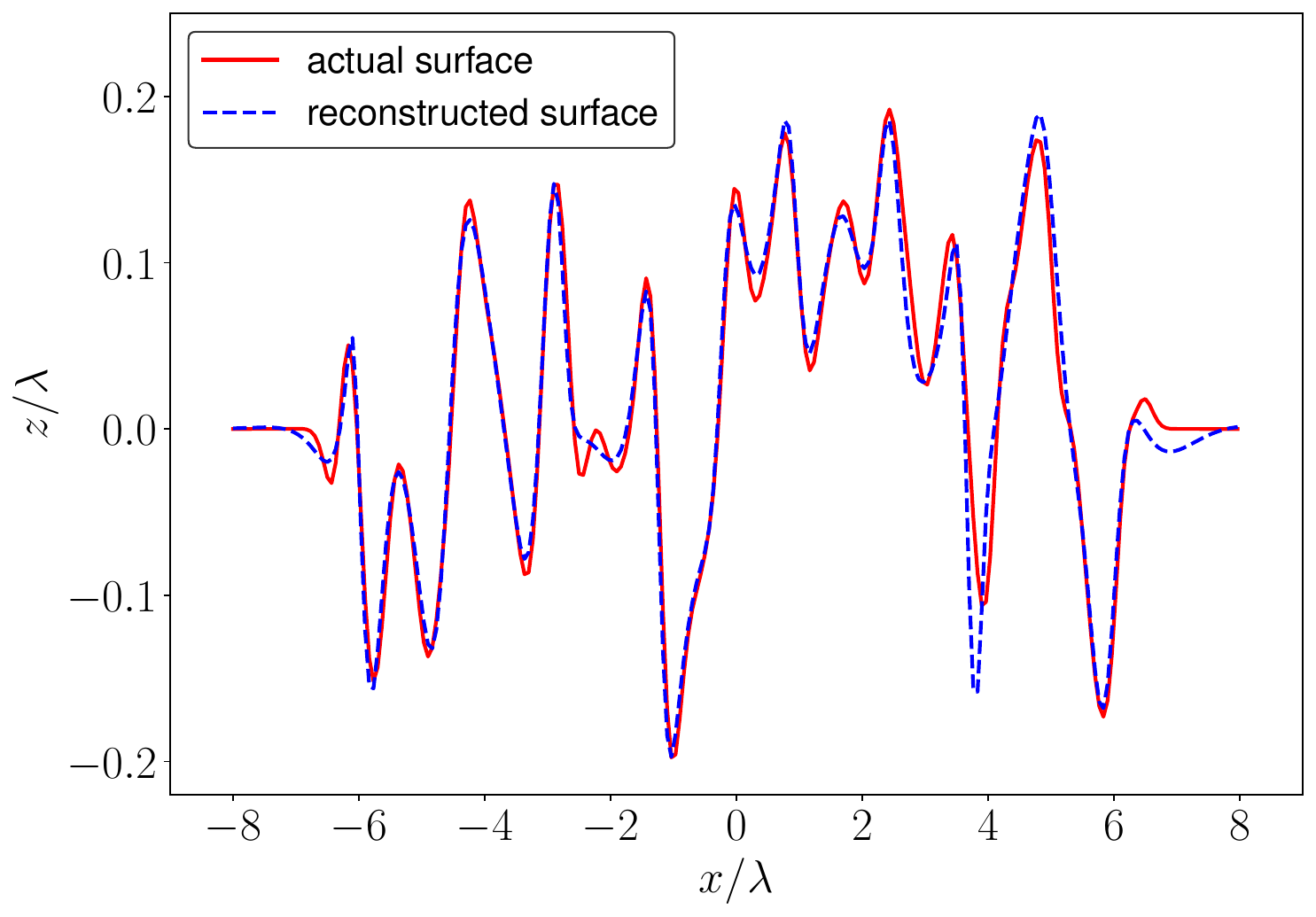}
\subcaption{case A, TM, $l=\frac{2}{5}\lambda$, $\Ni = 840$}
\end{subfigure}%
\vspace{0.22cm}
\begin{subfigure}{0.5\columnwidth}
\centering
\includegraphics[width=0.9\linewidth]{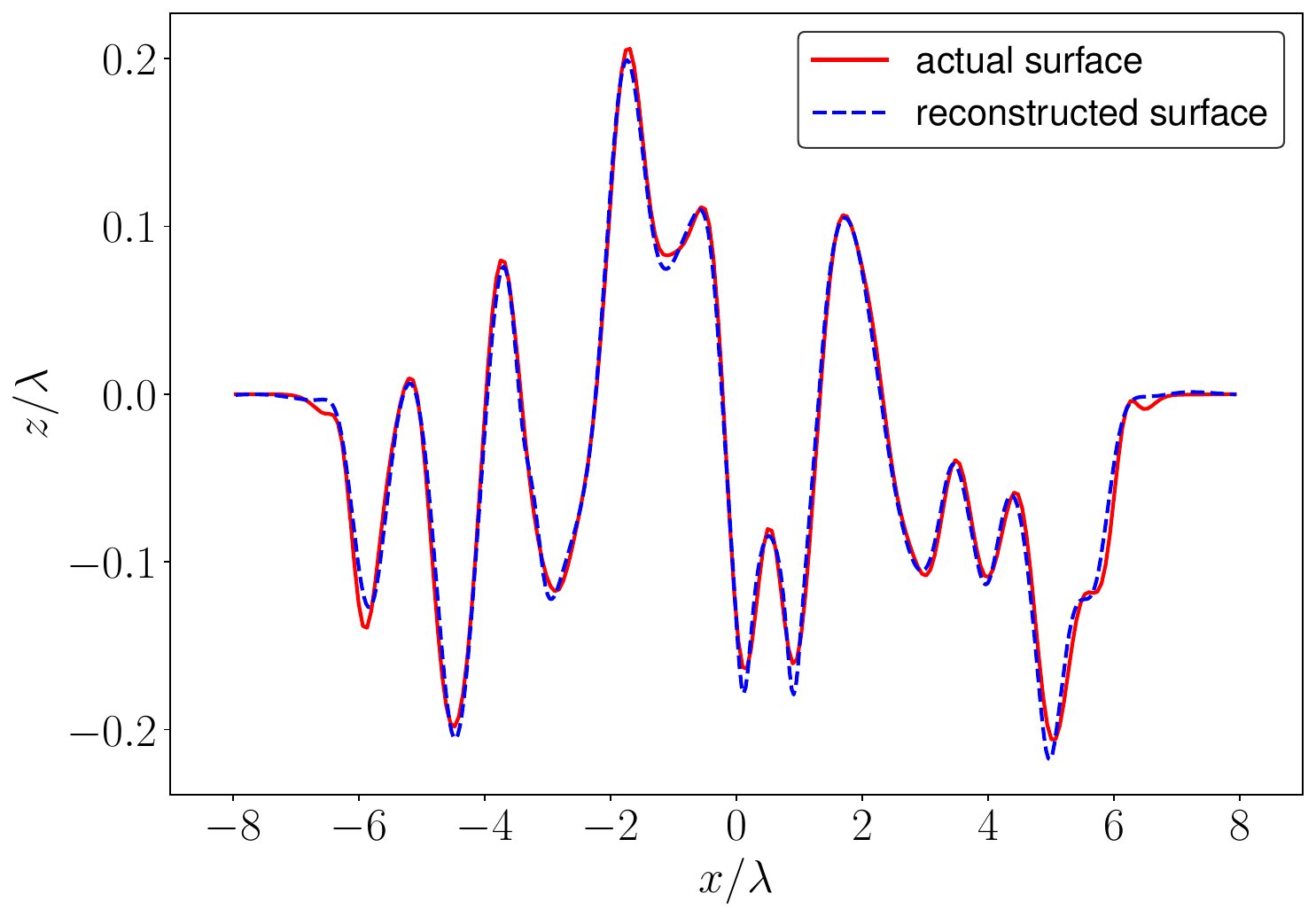}
\subcaption{case B, TM, $l=\frac{1}{2}\lambda$, $\Ni = 720$}
\end{subfigure}%
\begin{subfigure}{0.5\columnwidth}
\centering
\includegraphics[width=0.9\linewidth]{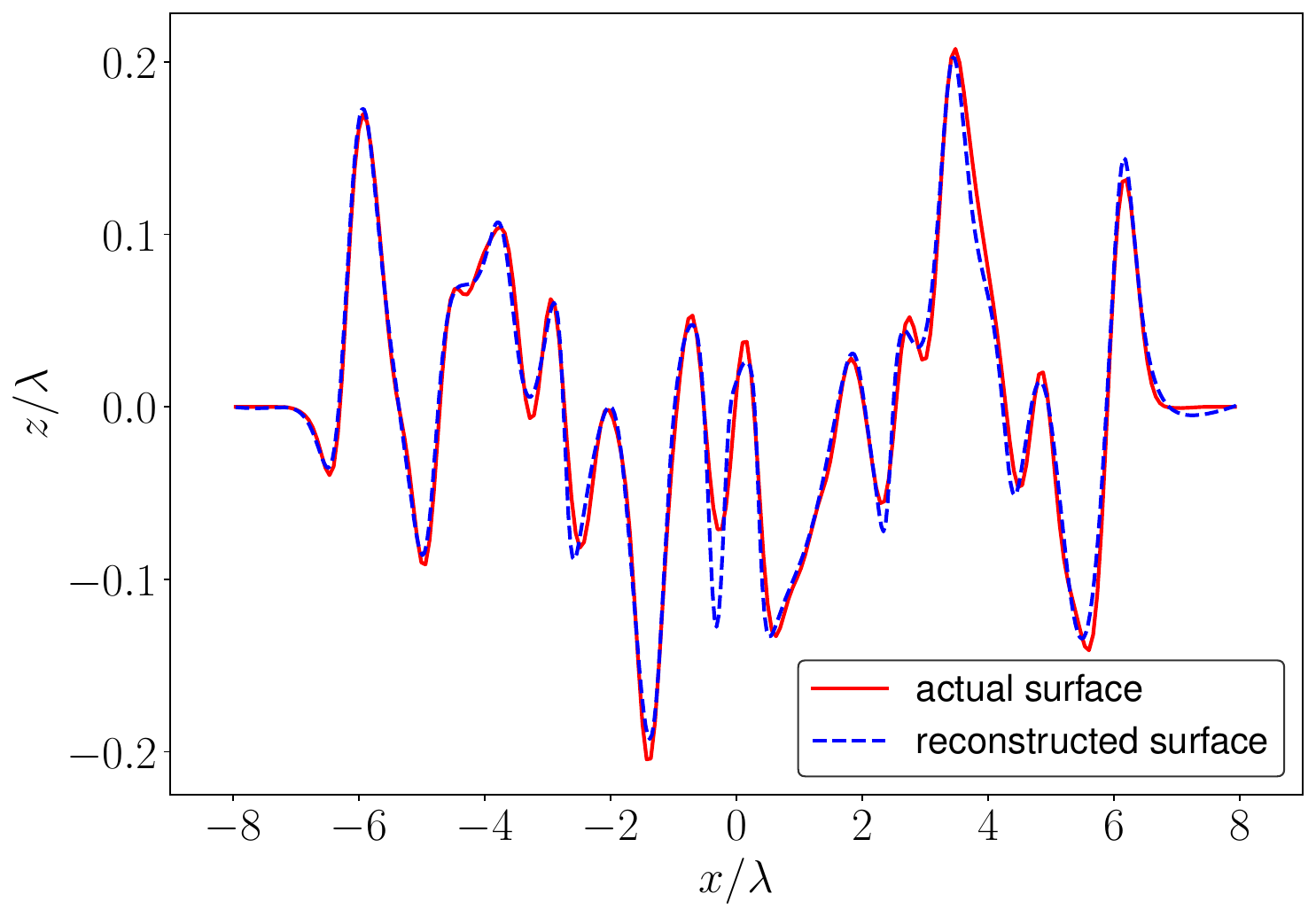}
\subcaption{case B, TE, $l=\frac{2}{5}\lambda$, $\Ni = 840$}
\end{subfigure}%
\caption{Rough surface reconstruction with respect to different values of
surface scales ($l$) using full scattered data (case A) with $10\%$ noise and
phaseless total field data (case B) with $3\%$ noise. The maximum number of
sampling points at each iteration $\Ni$ is also stated.}
\label{fig:ex3_recon_scale}
\end{figure}

\begin{figure}[t!]
  \centering
  \begin{subfigure}{0.5\columnwidth}
    \centering
    \includegraphics[width=0.9\linewidth]{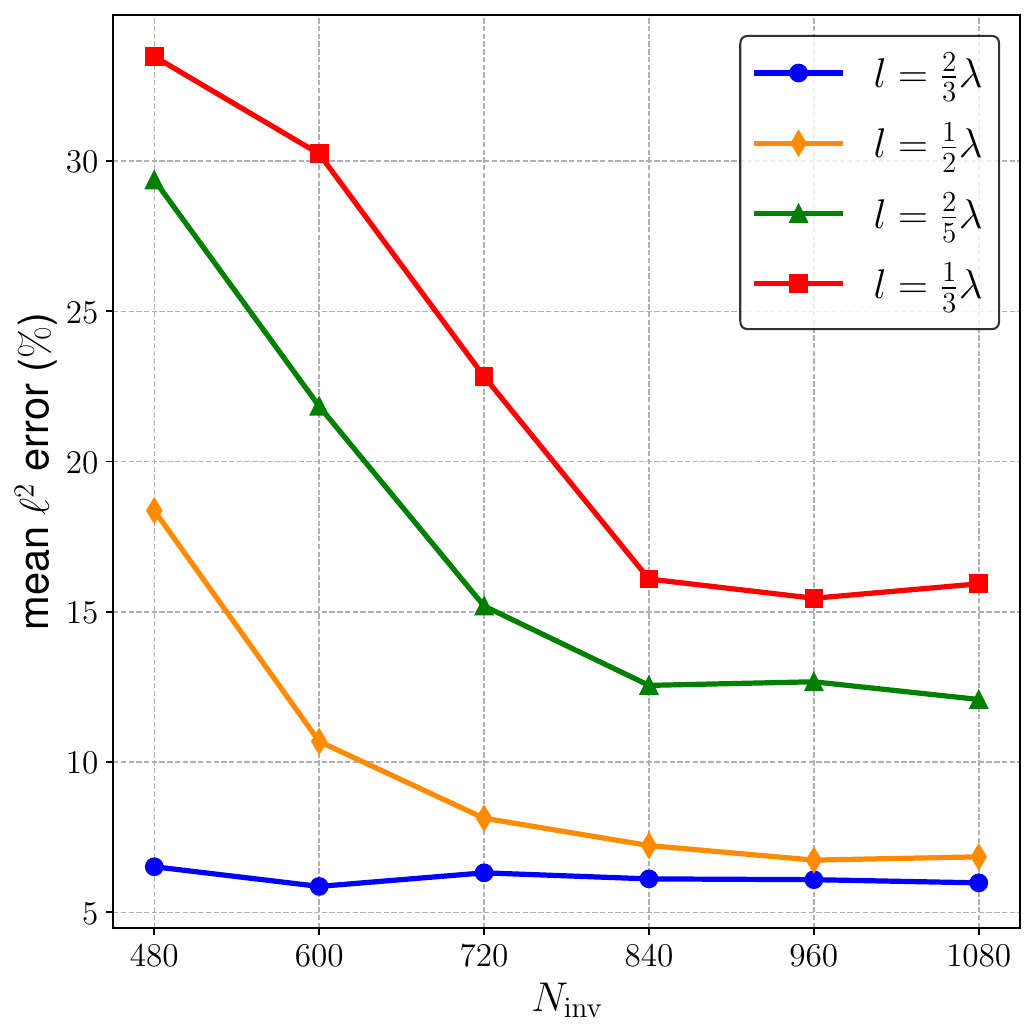}
    \subcaption{TE field, case A}
  \end{subfigure}%
  \begin{subfigure}{0.5\columnwidth}
    \centering
    \includegraphics[width=0.9\linewidth]{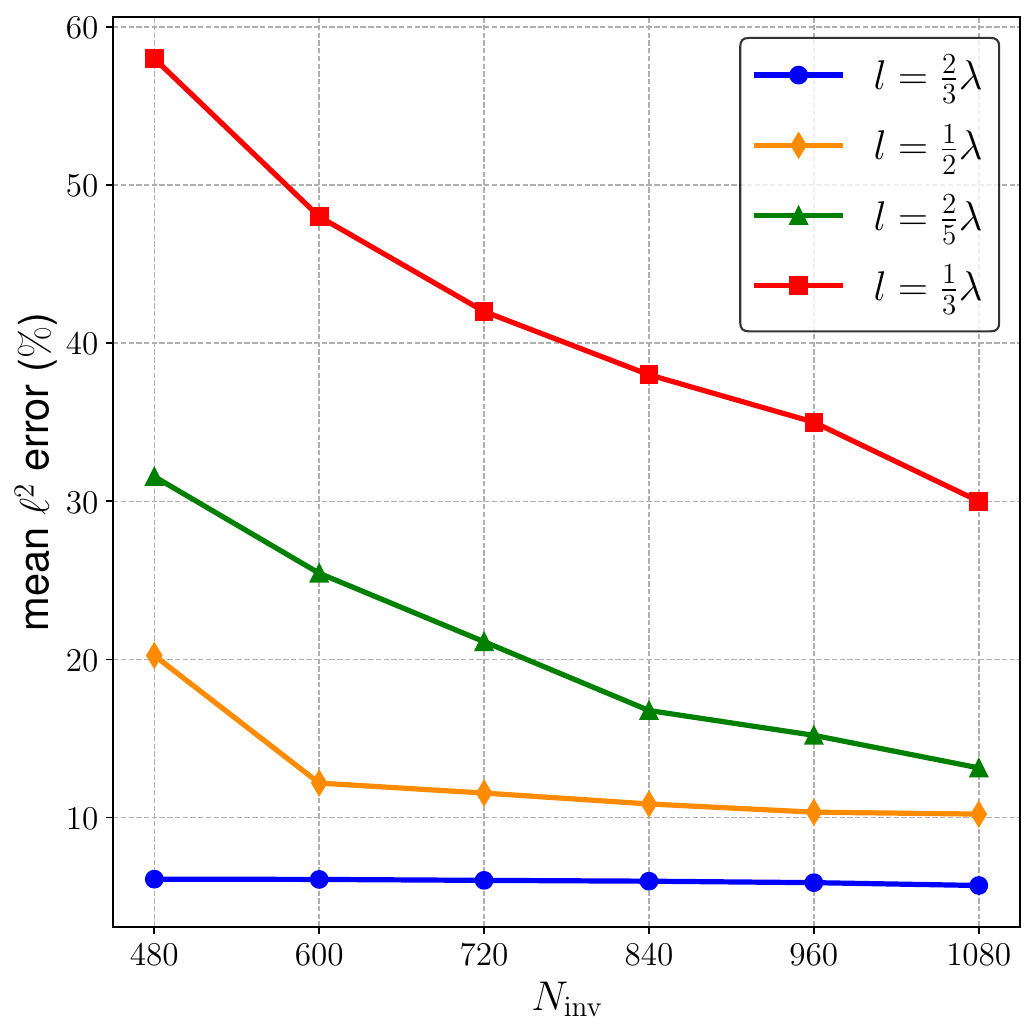}
    \subcaption{TM field, case B}
  \end{subfigure}
  \caption{The mean $\ell^2$-norm error obtained with different values of
  $\Ni$ (maximum number of testing points at each iteration) and different
  surface scales $l$.}
  \label{fig:ex3_scale_error}
\end{figure}

The surface scale ($l$ in \cref{eqn:acf}) controls the number of local
extrema in the rough surface; smaller $l$ gives rise to more oscillations. We
fix $h_{\max}- h_{\min} = 0.4 \lambda$ and $\zeta = 0.5 \lambda$. The number 
of observation points is still kept as $\Ni = 240$. As the surface scale
decreases, the value of $\Ni$ (maximum number of sampling points at each
iteration) has to be increased for two reasons: (i) it becomes more difficult
to reconstruct all the finer details for oscillatory surface and (ii) higher
resolution helps further reduce the loss during neural network training.
Some results of recovered surfaces with respect to relatively small surface
scales are shown in \cref{fig:ex3_recon_scale} for case A and case B, TE and
TM fields, and the values of $\Ni$ are also indicated.
With a larger value of $\Ni$, the recovered surface still captures the
overall shape of the original surface. As surface scale decreases,
discrepancies become more pronounced, particularly at the peaks and troughs
in the highly oscillatory region. \Cref{fig:ex3_scale_error} presents the mean
$\ell^2$-norm error obtained using different values of $\Ni$ with respect to
different surface scales ($l$).
It is clear that the error decreases with a larger value of $\Ni$.
Unsurprisingly, the error is small for less oscillatory surface (larger value
of $l$). The error remains steady provided that $\Ni$ is large enough for
$l = 2/3\lambda$ and $l = 1/2\lambda$.

\subsection{Reconstruction with respect to surface height}
\label{sec:example_height}

\begin{figure}
\captionsetup[subfigure]{justification=centering}
\centering
\begin{subfigure}{0.5\columnwidth}
\centering
\includegraphics[width=0.9\linewidth]{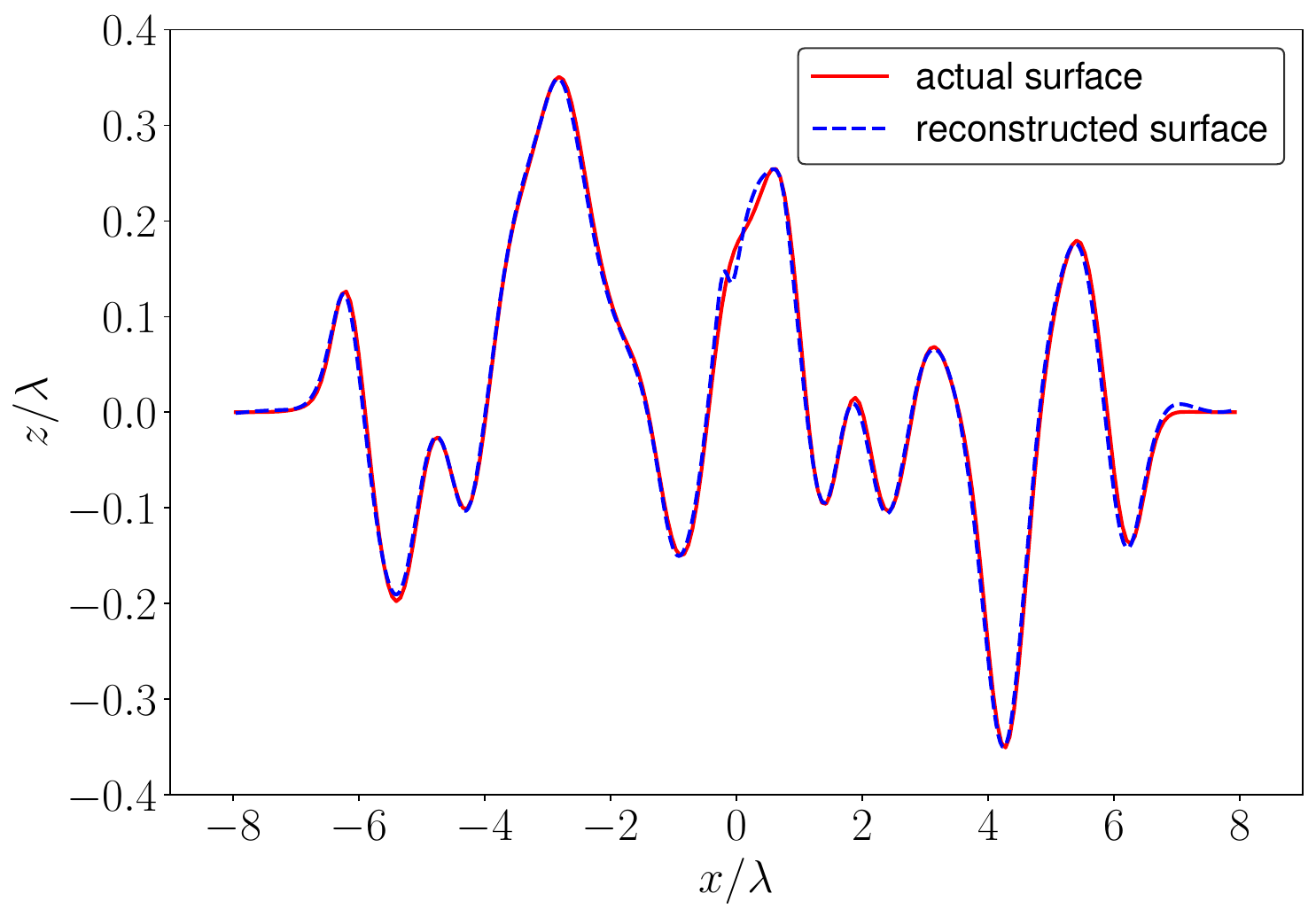}
\subcaption{case A, TM, \\ $h_{\max} - h_{\min} = 0.7\lambda$}
\end{subfigure}%
\begin{subfigure}{0.5\columnwidth}
\centering
\includegraphics[width=0.9\linewidth]{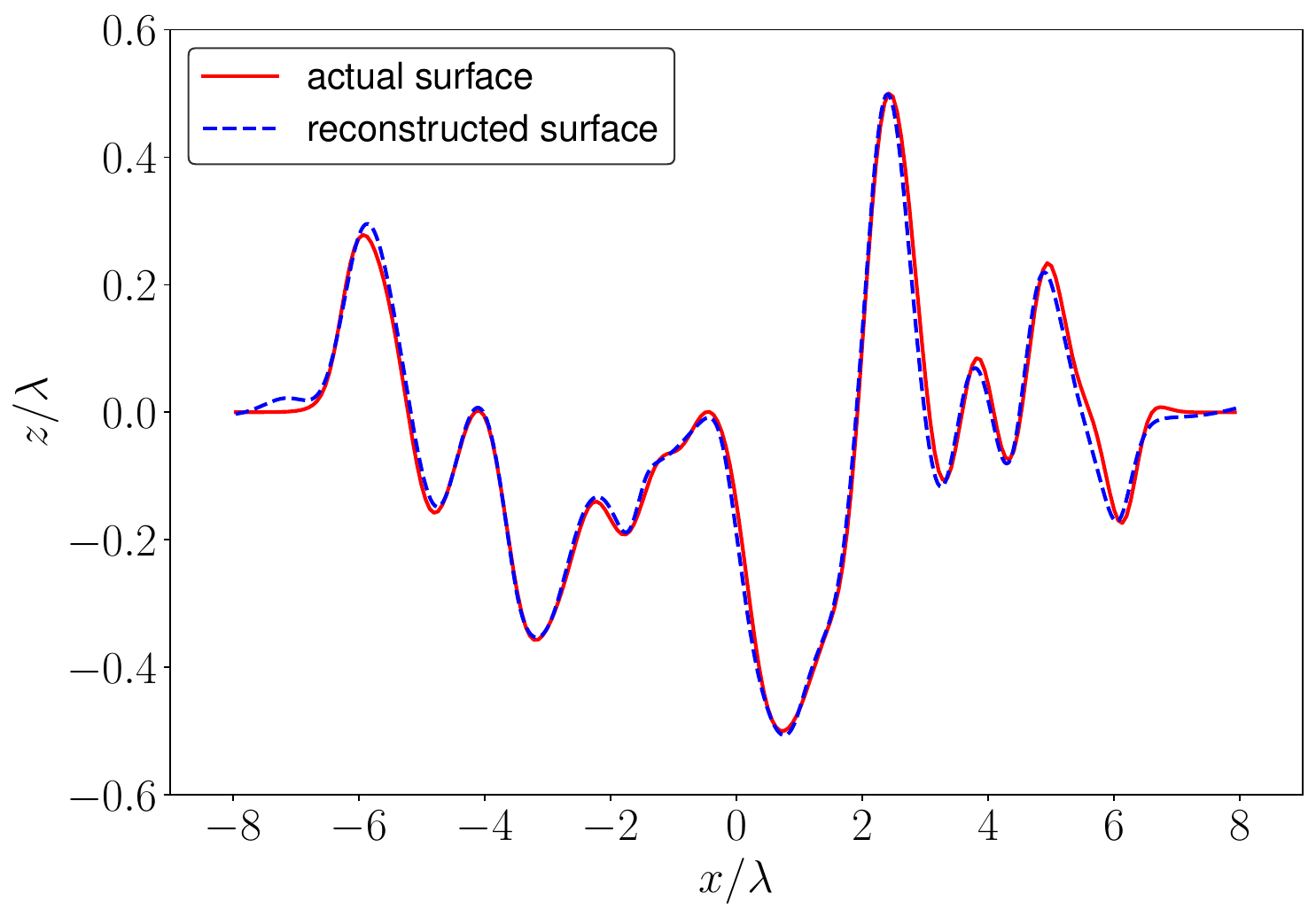}
\subcaption{case A, TE, \\ $h_{\max} - h_{\min} = \lambda$}
\end{subfigure}%
\hfill \hfill
\begin{subfigure}{0.5\columnwidth}
\centering
\includegraphics[width=0.9\linewidth]{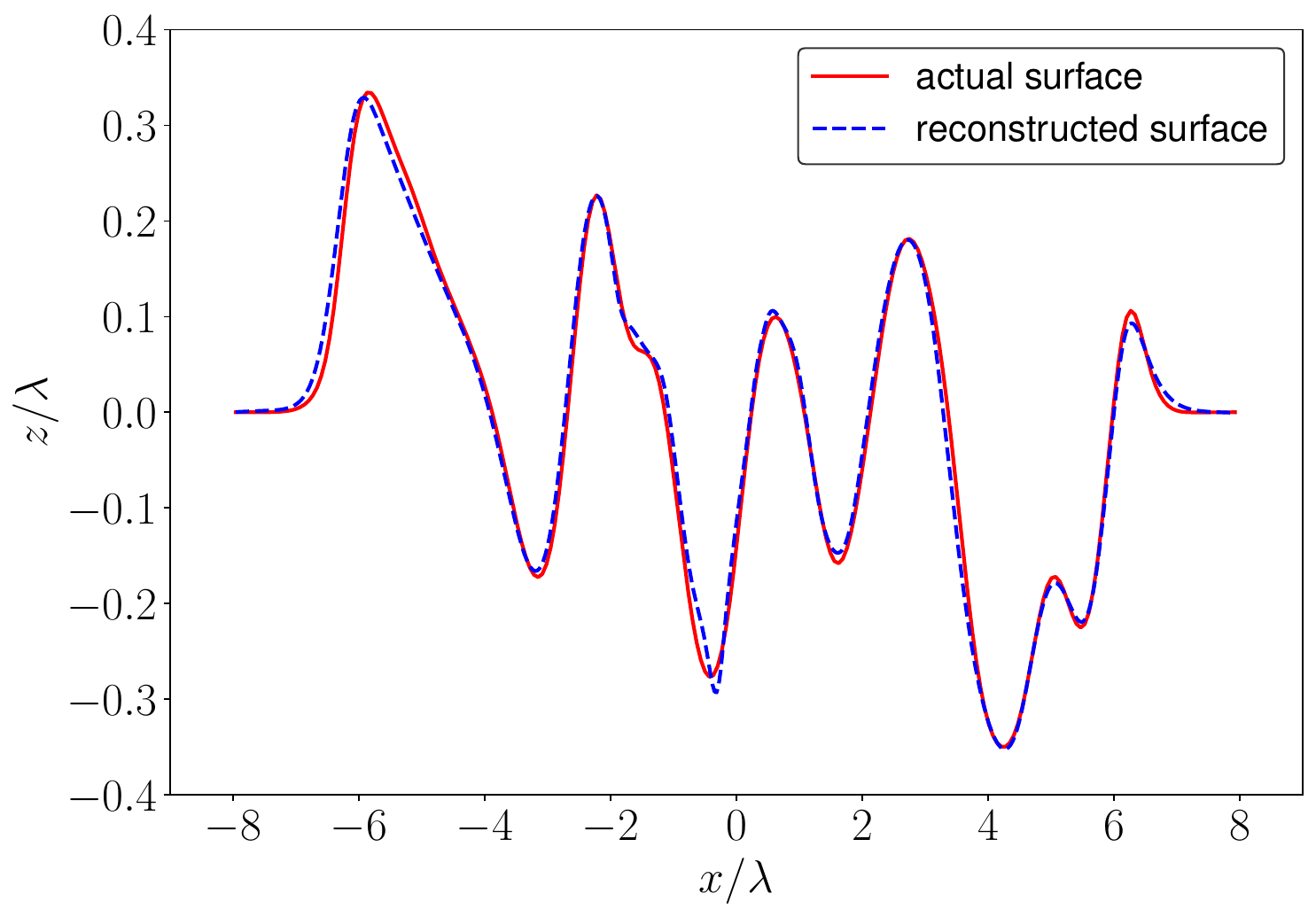}
\subcaption{case B, TE, \\ $h_{\max} - h_{\min} = 0.7\lambda$}
\end{subfigure}%
\begin{subfigure}{0.5\columnwidth}
\centering
\includegraphics[width=0.9\linewidth]{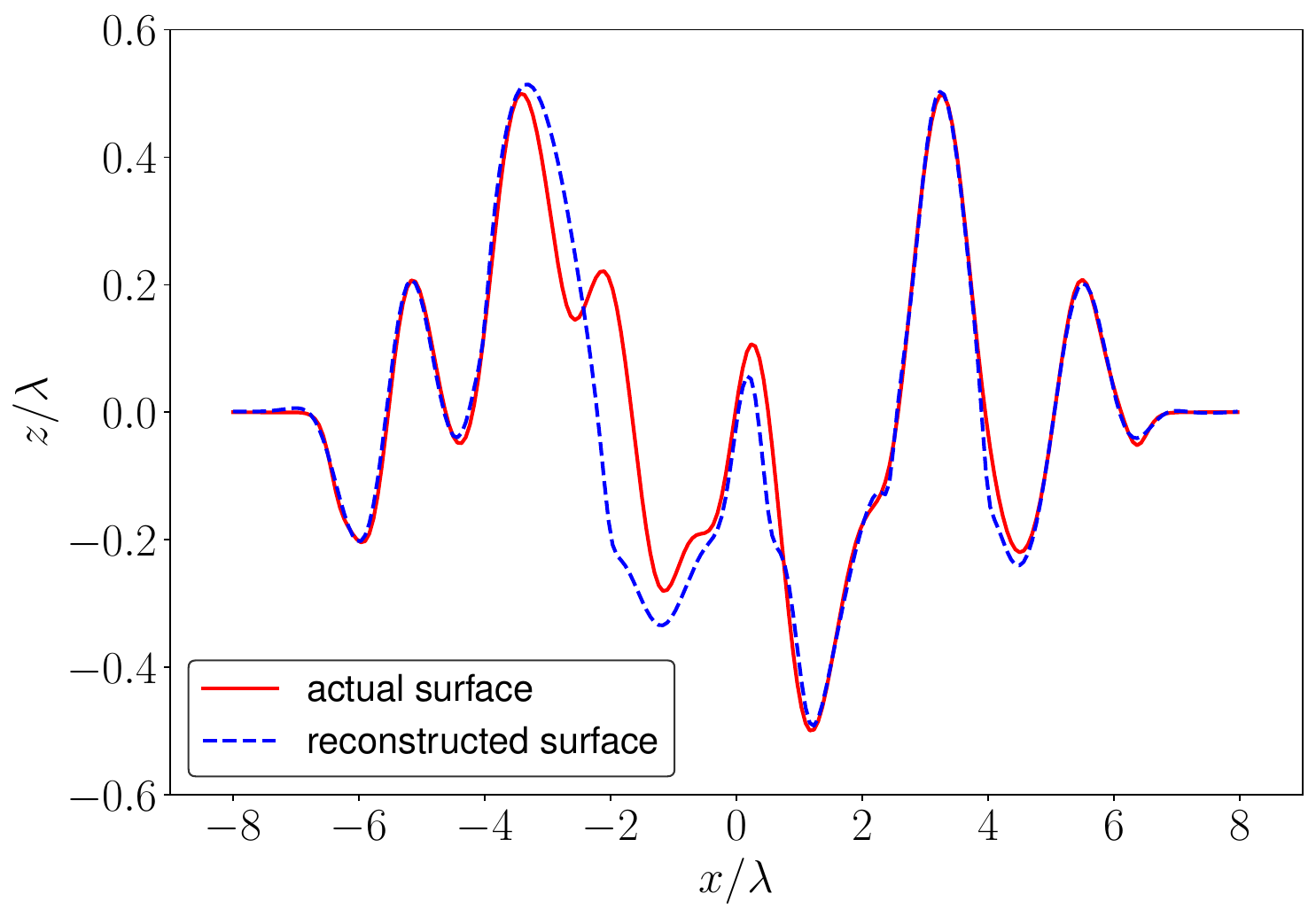}
\subcaption{case B, TM, \\ $h_{\max} - h_{\min} = \lambda$}
\end{subfigure}%
\caption{Reconstruction of rough surfaces compared to actual surfaces with
respect to different peak-to-trough heights ($h_{\max}-h_{\min}$) for case A
(full scattered data) with $10\%$ noise and case B (phaseless total field
data) with $3\%$ noise.}
\label{fig:ex4_height_recons}
\end{figure}

Surface height, in particular the peak-to-trough height
($h_{\max} -h_{\min}$), is a key factor in the inverse problem. Increased
height from peak to trough typically leads to stronger surface scattering,
resulting in degraded performance of reconstruction algorithms. Here, we
solve the problem with parameters of $l = 2/3\lambda$, $\zeta = 2.5h_{\max}$,
$\Nd = 240$ and $\Ni = 480$. Some reconstructions of rough surfaces with
large peak-to-trough height values for TE and TM fields in case A and case B
are shown in \cref{fig:ex4_height_recons}. The performance degrades for
surfaces with larger peak-to-trough height. Overall, despite variations, all
the recovered surfaces still manage to preserve the essential feature and
shape of original surfaces. This demonstrates the robustness of the method when dealing with surface height variations.
The mean $\ell^2$-norm error together with the standard deviation with
respect to different values of $h_{\max} - h_{\min}$ is presented in
\cref{table:error_height} for case A data with $10\%$ noise and case B data
with $3\%$ noise. There is an obvious increase in error for larger surface
height, as well as the standard deviation. However, the error is still
considerably small when the peak-to-trough height is less than $\lambda$.
\setlength{\tabcolsep}{4.5pt}
\renewcommand{\arraystretch}{1.3}
\begin{table}[t]
\centering
\caption{Mean error in $\ell^2$-norm (\%) and standard deviation 
between the actual and recovered surfaces with respect to different
peak-to-trough heights}
\label{table:error_height}
\begin{tabular}{l|cc|cc}
\hline
\multirow{2}{*}{$h_{\max}$$-$$h_{\min}$} & 
\multicolumn{2}{c|}{case A} & \multicolumn{2}{c}{case B}   \\ 
& TE           & TM           & TE           & TM          \\ 
\hline
$0.4\lambda$  & 6.57$\pm$0.84  & 5.87$\pm$0.67  & 8.52$\pm$1.01   
& 10.72$\pm$2.24  \\
$0.6\lambda$  & 7.13$\pm$0.95  & 6.47$\pm$0.83  & 10.26$\pm$0.96
& 11.38$\pm$2.87   \\
$0.8\lambda$  & 7.39$\pm$1.12  & 8.91$\pm$1.28  & 11.04$\pm$1.54
& 11.57$\pm$2.79 \\
$1.0\lambda$  & 10.55$\pm$2.43  & 10.36$\pm$2.60  & 11.37$\pm$2.77
& 12.45$\pm$3.51   \\
$1.2\lambda$  & 15.20$\pm$3.11  & 17.49$\pm$3.52  & 28.96$\pm$5.84
& 36.08$\pm$12.39  \\
\hline
\end{tabular}
\end{table}
%

\subsection{Reconstruction with respect to incident wave}
\label{sec:example_wavenumber}

\begin{figure}[t]
\centering
\begin{subfigure}{0.5\columnwidth}
\centering
\includegraphics[width=0.9\linewidth]{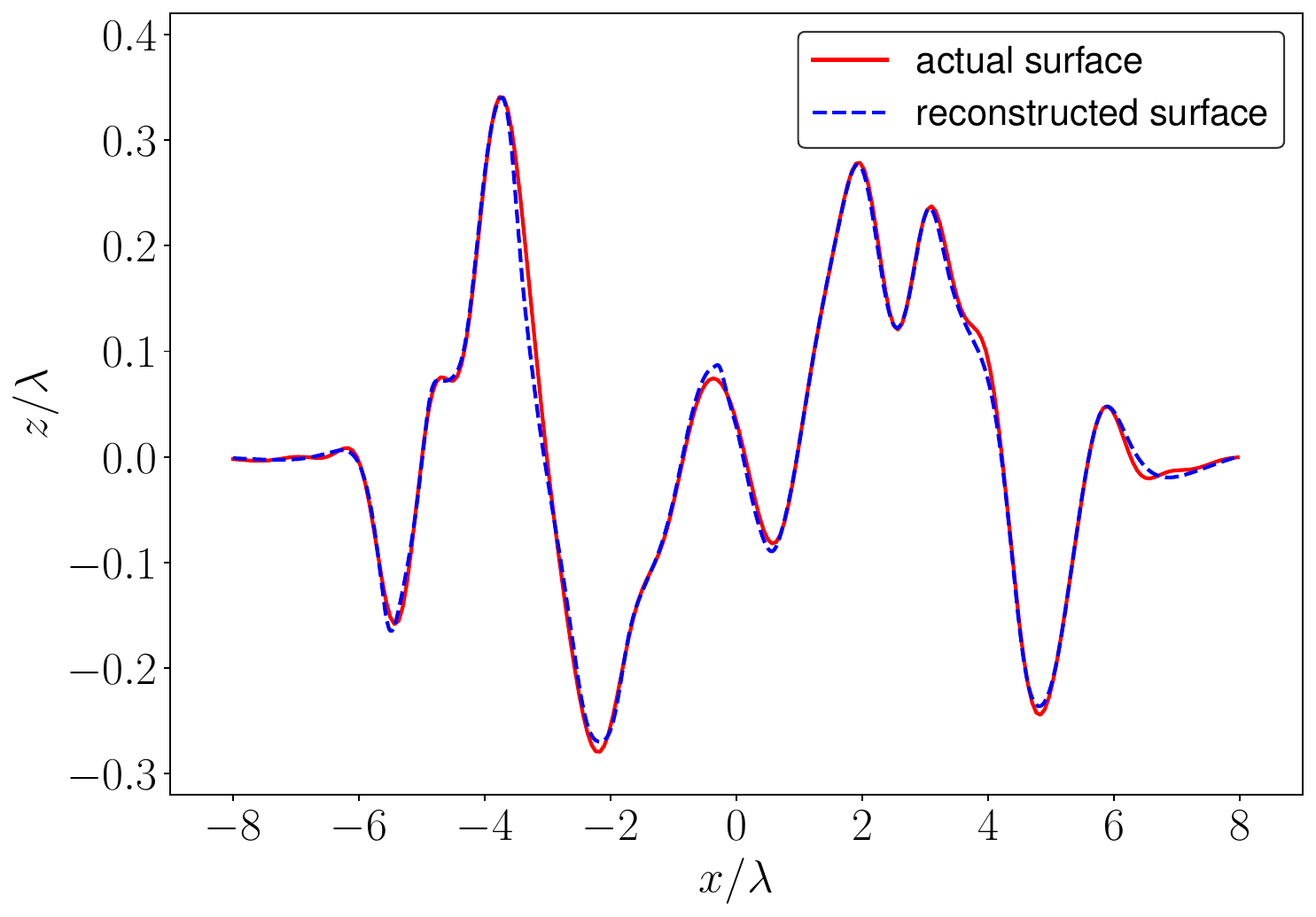}
\subcaption{TM field, case A}
\end{subfigure}%
\begin{subfigure}{0.5\columnwidth}
\centering
\includegraphics[width=0.9\linewidth]{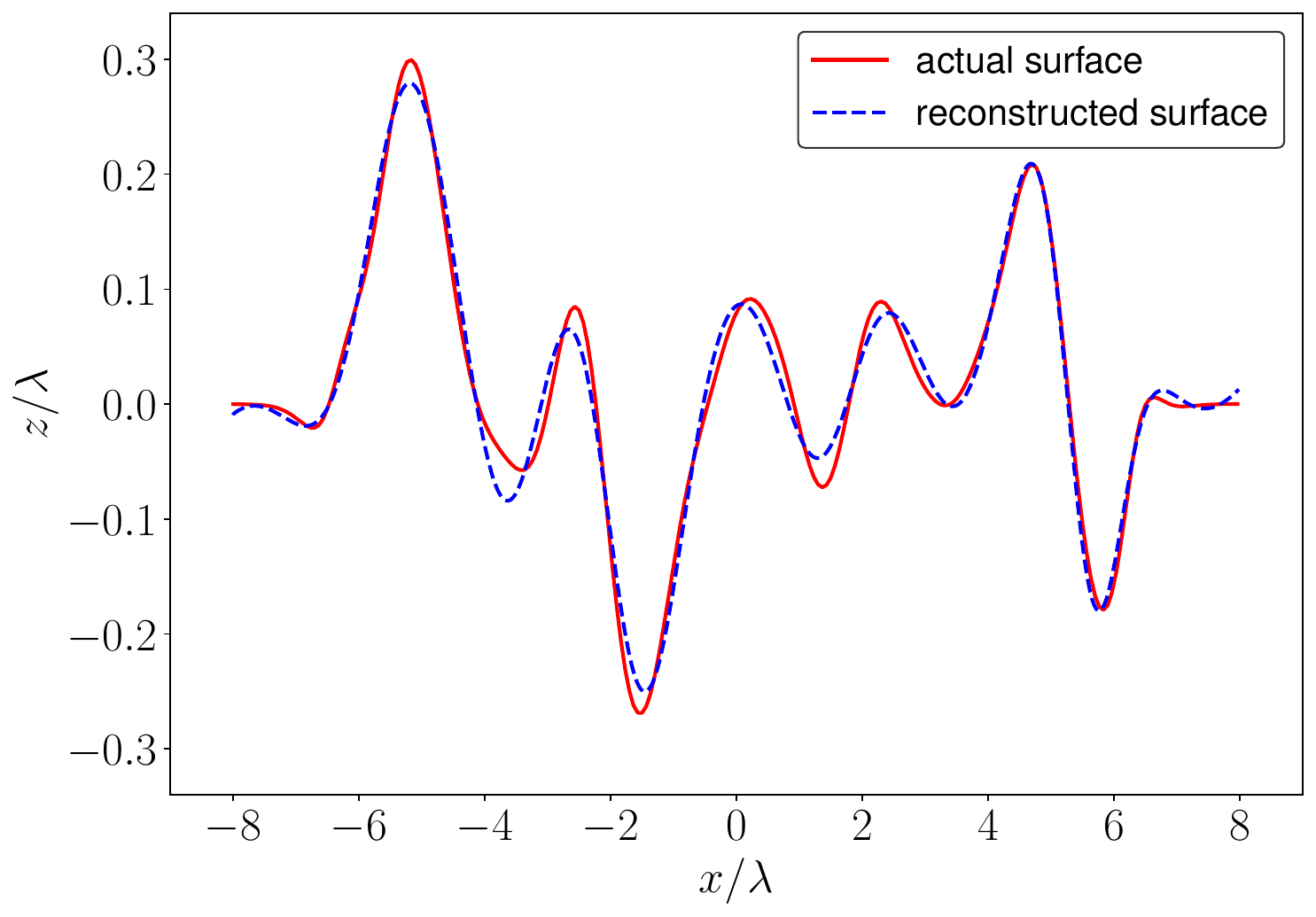}
\subcaption{TE field, case B}
\end{subfigure}%
\caption{Reconstruction of rough surfaces compared to actual surfaces from
the plane wave of $k =6.67 \pi$ and $\alpha = -10 / \pi$ for case A (full
scattered data) with $10\%$ noise and case B (phaseless total field data)
with $3\%$ noise.}
\label{fig:ex5_inc_recons}
\end{figure}

An important aspect of this method is its flexibility with respect to the
incident field (wavenumber and angle of incidence). We consider the problem
setting of $l=2/3 \lambda$, $h_{\max} - h_{\min} = 0.6 \lambda$,
$\zeta = 0.6 \lambda$, $\Nd = 240$, and $\Ni = 480$, while the wavenumber has 
been increased to  $k = 6.67 \pi$ (corresponding to frequency of 1GHz and
wavelength of $0.3$ meters) and the angle of grazing is set to be $-\pi/9$
(corresponding to $10^{\circ}$). The results obtained using the same
structure of neural networks (\cref{table:structure}) for case A TM data with
$10\%$ noise and case B TE data with $3\%$ noise are shown in
\cref{fig:ex5_inc_recons}.
The reconstructed surface still closely matches the actual surface, which
validates the robustness of the method with respect to incident wave
characteristics. We make a remark here that the error increases when using incident field with high wavenumber (frequency). However the error can be 
reduced via employing deeper neural networks.

\section{Conclusions}
\label{sec:conclusions}

A deep learning approach based on physics--informed neural networks (PINN)
has been developed to recover profile of rough surfaces from observed
electromagnetic field data. In this approach, the unknown surface height is
approximated by a neural network. Thy `physical information' (equations)
used in the neural network are the integral equations relating the surface,
incident and scattered field. With the predicted surface (neural network) and
its spatial derivatives (automatic differentiation), we calculate the
scattered field via the method of moments (MOM). The neural network is then
trained via optimizing the loss between the observed and calculated field
data. It was shown that the method can accurately recover highly varying
random rough surfaces.

A particular aspect of this approach is that it is an unsupervised
reconstruction method, namely, it does not require any `\emph{a priori}'
`solution' data. This is particularly appealing in the rough surface
reconstruction field, where it is usually difficult and time-consuming to
generate surface data in practice, rather a more typical inverse problem is
to recover specific unknown surfaces from the measured field data. The
proposed method, therefore, opens up the possibility treating real--world
applications through a fast and robust deep learning approach,
such as non--destructive testing and microwave remote sensing.

Two types of data were considered in this paper: full scattered field data
and phaseless total field data, which results from both TE and TM incident
fields. The performance of the method is validated by a large number of test
cases with Gaussian-correlated random rough surfaces. The method has also
shown a strong robustness with respect to a range of different regimes,
including noisy data, surface scale (degree of surface oscillations), surface
height (peak-to-trough), and incident wave (wavenumber and incidence angle).

It is noted that even though this paper shows the results of rough surface
reconstruction in 2D, the approach can be easily extended to 3D problems. For
3D applications, the main difficulty becomes the time complexity of method of
moments, which involves a step of inverting a linear system of size
$N^2 \times N^2$, where $N$ is the discretization size in one axis. Another
important problem to which this method can extend straightforwardly is the
design of inverse waveguide problem. In 2D version of inverse waveguide
problem, a pair of rough surfaces need to be reconstructed with respect to a
set of field data.

\section{ACKNOWLEDGEMENT}

YC was supported by the Youth Program of the Natural Science Foundation
of Jiangsu Province (No.~BK20230466), the Jiangsu Funding Program for
Excellent Postdoctoral Talent (No.~2022ZB584), and Jiangsu Shuangchuang
Project (JSSCTD202209). CW was supported by the National Natural Science 
Foundation of China under Grants No. 62301532, in part by the Natural Science 
Foundation of Jiangsu Province under Grant No. BK20230282

\bibliographystyle{IEEEtran}
\bibliography{references}

\begin{thebibliography}{10}
\providecommand{\url}[1]{#1}
\csname url@samestyle\endcsname
\providecommand{\newblock}{\relax}
\providecommand{\bibinfo}[2]{#2}
\providecommand{\BIBentrySTDinterwordspacing}{\spaceskip=0pt\relax}
\providecommand{\BIBentryALTinterwordstretchfactor}{4}
\providecommand{\BIBentryALTinterwordspacing}{\spaceskip=\fontdimen2\font plus
\BIBentryALTinterwordstretchfactor\fontdimen3\font minus
  \fontdimen4\font\relax}
\providecommand{\BIBforeignlanguage}[2]{{%
\expandafter\ifx\csname l@#1\endcsname\relax
\typeout{** WARNING: IEEEtran.bst: No hyphenation pattern has been}%
\typeout{** loaded for the language `#1'. Using the pattern for}%
\typeout{** the default language instead.}%
\else
\language=\csname l@#1\endcsname
\fi
#2}}
\providecommand{\BIBdecl}{\relax}
\BIBdecl

\bibitem{retov}
A.~Schatzberg and A.~J. Devaney, ``Rough surface inverse scattering within the
  {R}ytov approximation,'' \emph{Journal of the Optical Society of America A},
  vol.~10, no.~5, pp. 942--950, 1993.

\bibitem{chen2018rough1}
Y.~Chen and M.~Spivack, ``Rough surface reconstruction at grazing angles by an
  iterated marching method,'' \emph{Journal of the Optical Society of America
  A}, vol.~35, no.~4, pp. 504--513, 2018.

\bibitem{chen2018recovery}
Y.~Chen, O.~Rath~Spivack, and M.~Spivack, ``{Recovery of rough surface in
  ducting medium from grazing angle scattered wave},'' \emph{Journal of Applied
  Physics}, vol. 124, no.~8, p. 084901, 08 2018.

\bibitem{DESTOUCHES2001233}
N.~Destouches, C.-A. Guérin, M.~Lequime, and H.~Giovannini, ``Determination of
  the phase of the diffracted field in the optical domain: Application to the
  reconstruction of surface profiles,'' \emph{Optics Communications}, vol. 198,
  no.~4, pp. 233--239, 2001.

\bibitem{Akduman_2006}
I.~Akduman, R.~Kress, and A.~Yapar, ``Iterative reconstruction of dielectric
  rough surface profiles at fixed frequency,'' \emph{Inverse Problems},
  vol.~22, no.~3, p. 939, may 2006.

\bibitem{newton1}
R.~Kress and T.~Tran, ``Inverse scattering for a locally perturbed
  half-plane,'' \emph{Inverse Problems}, vol.~16, no.~5, p. 1541, oct 2000.

\bibitem{newton2}
A.~Yapar, O.~Ozdemir, H.~Sahinturk, and I.~Akduman, ``A {N}ewton method for the
  reconstruction of perfectly conducting slightly rough surface profiles,''
  \emph{IEEE Transactions on Antennas and Propagation}, vol.~54, no.~1, pp.
  275--279, 2006.

\bibitem{newton3}
P.~Mojabi and J.~LoVetri, ``Overview and classification of some regularization
  techniques for the {Gauss-Newton} inversion method applied to inverse
  scattering problems,'' \emph{IEEE Transactions on Antennas and Propagation},
  vol.~57, no.~9, pp. 2658--2665, 2009.

\bibitem{newton4}
G.~Bozza and M.~Pastorino, ``An inexact {N}ewton-based approach to microwave
  imaging within the contrast source formulation,'' \emph{IEEE Transactions on
  Antennas and Propagation}, vol.~57, no.~4, pp. 1122--1132, 2009.

\bibitem{sampling1}
J.~Li, J.~Yang, and B.~Zhang, ``A linear sampling method for inverse acoustic
  scattering by a locally rough interface,'' \emph{Inverse Problems and
  Imaging}, vol.~15, no.~5, pp. 1247--1267, 2021.

\bibitem{sampling2}
X.~Xu, B.~Zhang, and H.~Zhang, ``Uniqueness and direct imaging method for
  inverse scattering by locally rough surfaces with phaseless near-field
  data,'' \emph{SIAM Journal on Imaging Sciences}, vol.~12, no.~1, pp.
  119--152, 2019.

\bibitem{sampling3}
X.~Ji, X.~Liu, and B.~Zhang, ``Inverse acoustic scattering with phaseless far
  field data: Uniqueness, phase retrieval, and direct sampling methods,''
  \emph{SIAM Journal on Imaging Sciences}, vol.~12, no.~2, pp. 1163--1189,
  2019.

\bibitem{bao2016shape}
G.~Bao and L.~Zhang, ``Shape reconstruction of the multi-scale rough surface
  from multi-frequency phaseless data,'' \emph{Inverse Problems}, vol.~32,
  no.~8, p. 085002, 2016.

\bibitem{dolcetti2021robust}
G.~Dolcetti, M.~Alkmim, J.~Cuenca, L.~{De Ryck}, and A.~Krynkin, ``Robust
  reconstruction of scattering surfaces using a linear microphone array,''
  \emph{Journal of Sound and Vibration}, vol. 494, p. 115902, 2021.

\bibitem{sefer1}
A.~Sefer, ``Locally perturbed inaccessible rough surface profile reconstruction
  via phaseless scattered field data,'' \emph{IEEE Transactions on Geoscience
  and Remote Sensing}, vol.~60, pp. 1--8, 2022.

\bibitem{sefer2}
A.~Sefer and A.~Yapar, ``Inverse scattering by perfectly electric conducting
  ({PEC}) rough surfaces: An equivalent model with line sources,'' \emph{IEEE
  Transactions on Geoscience and Remote Sensing}, vol.~60, pp. 1--9, 2022.

\bibitem{sefer3}
A.~Sefer, A.~Yapar, and T.~Yelkenci, ``Imaging of rough surfaces by {RTM}
  method,'' \emph{IEEE Transactions on Geoscience and Remote Sensing}, vol.~62,
  pp. 1--12, 2024.

\bibitem{iterative1}
A.~Sefer and A.~Yapar, ``An iterative algorithm for imaging of rough surfaces
  separating two dielectric media,'' \emph{IEEE Transactions on Geoscience and
  Remote Sensing}, vol.~59, no.~2, pp. 1041--1051, 2021.

\bibitem{chen2018rough2}
Y.~Chen, O.~R. Spivack, and M.~Spivack, ``Rough surface reconstruction from
  phaseless single frequency data at grazing angles,'' \emph{Inverse Problems},
  vol.~34, no.~12, p. 124002, 2018.

\bibitem{iterative3}
F.~Qu, B.~Zhang, and H.~Zhang, ``A novel integral equation for scattering by
  locally rough surfaces and application to the inverse problem: The {N}eumann
  case,'' \emph{SIAM Journal on Scientific Computing}, vol.~41, no.~6, pp.
  A3673--A3702, 2019.

\bibitem{wombell1991reconstruction}
R.~J. Wombell and J.~A. DeSanto, ``Reconstruction of rough-surface profiles
  with the {K}irchhoff approximation,'' \emph{Journal of the Optical Society of
  America A}, vol.~8, no.~12, pp. 1892--1897, Dec 1991.

\bibitem{dl1}
Y.~Zhou, Y.~Zhong, Z.~Wei, T.~Yin, and X.~Chen, ``An improved deep learning
  scheme for solving 2-{D} and 3-{D} inverse scattering problems,'' \emph{IEEE
  Transactions on Antennas and Propagation}, vol.~69, no.~5, pp. 2853--2863,
  2020.

\bibitem{dl2}
K.~Xu, L.~Wu, X.~Ye, and X.~Chen, ``Deep learning-based inversion methods for
  solving inverse scattering problems with phaseless data,'' \emph{IEEE
  Transactions on Antennas and Propagation}, vol.~68, no.~11, pp. 7457--7470,
  2020.

\bibitem{dl3}
K.~Xu, C.~Zhang, X.~Ye, and R.~Song, ``Fast full-wave electromagnetic inverse
  scattering based on scalable cascaded convolutional neural networks,''
  \emph{IEEE Transactions on Geoscience and Remote Sensing}, vol.~60, pp.
  1--11, 2022.

\bibitem{dl4}
R.~Guo, X.~Song, M.~Li, F.~Yang, S.~Xu, and A.~Abubakar, ``Supervised descent
  learning technique for 2-{D} microwave imaging,'' \emph{IEEE Transactions on
  Antennas and Propagation}, vol.~67, no.~5, pp. 3550--3554, 2019.

\bibitem{dl5}
M.~Sabbaghi, J.~Zhang, and G.~W. Hanson, ``Machine learning target count
  prediction in electromagnetics using neural networks,'' \emph{IEEE
  Transactions on Antennas and Propagation}, vol.~70, no.~8, pp. 6171--6183,
  2022.

\bibitem{dl6}
Y.~Sanghvi, Y.~Kalepu, and U.~K. Khankhoje, ``Embedding deep learning in
  inverse scattering problems,'' \emph{IEEE Transactions on Computational
  Imaging}, vol.~6, pp. 46--56, 2020.

\bibitem{dl7}
G.~Chen, P.~Shah, J.~Stang, and M.~Moghaddam, ``Learning-assisted multimodality
  dielectric imaging,'' \emph{IEEE Transactions on Antennas and Propagation},
  vol.~68, no.~3, pp. 2356--2369, 2020.

\bibitem{dl8}
L.~Li, L.~G. Wang, F.~L. Teixeira, C.~Liu, A.~Nehorai, and T.~J. Cui,
  ``{DeepNIS}: Deep neural network for nonlinear electromagnetic inverse
  scattering,'' \emph{IEEE Transactions on Antennas and Propagation}, vol.~67,
  no.~3, pp. 1819--1825, 2019.

\bibitem{dl9}
Z.~Zong, Y.~Wang, and Z.~Wei, ``A wavelet-based compressive deep learning
  scheme for inverse scattering problems,'' \emph{IEEE Transactions on
  Geoscience and Remote Sensing}, vol.~60, pp. 1--11, 2022.

\bibitem{cnn1}
I.~Aydin, G.~Budak, A.~Sefer, and A.~Yapar, ``{CNN}-based deep learning
  architecture for electromagnetic imaging of rough surface profiles,''
  \emph{IEEE Transactions on Antennas and Propagation}, vol.~70, no.~10, pp.
  9752--9763, 2022.

\bibitem{cnn2}
A.~S. Izde~Aydin, Guven~Budak and A.~Yapar, ``Recovery of impenetrable rough
  surface profiles via {CNN}-based deep learning architecture,''
  \emph{International Journal of Remote Sensing}, vol.~43, no. 15-16, pp.
  5658--5685, 2022.

\bibitem{pinn1}
M.~Raissi, P.~Perdikaris, and G.~Karniadakis, ``Physics--informed neural
  networks: A deep learning framework for solving forward and inverse problems
  involving nonlinear partial differential equations,'' \emph{Journal of
  Computational Physics}, vol. 378, pp. 686--707, 2019.

\bibitem{pinn2}
G.~E. Karniadakis, I.~G. Kevrekidis, L.~Lu, P.~Perdikaris, S.~Wang, and
  L.~Yang, ``Physics--informed machine learning,'' \emph{Nature Reviews
  Physics}, vol.~3, no.~6, pp. 422--440, 2021.

\bibitem{pinn_direct1}
M.~Rasht-Behesht, C.~Huber, K.~Shukla, and G.~E. Karniadakis,
  ``Physics-informed neural networks ({PINNs}) for wave propagation and full
  waveform inversions,'' \emph{Journal of Geophysical Research: Solid Earth},
  vol. 127, no.~5, p. e2021JB023120, 2022.

\bibitem{pinn_direct2}
S.~Alkhadhr, X.~Liu, and M.~Almekkawy, ``Modeling of the forward wave
  propagation using physics-informed neural networks,'' in \emph{2021 IEEE
  International Ultrasonics Symposium (IUS)}, 2021, pp. 1--4.

\bibitem{pinn_direct3}
Z.~Yin, G.-Y. Li, Z.~Zhang, Y.~Zheng, and Y.~Cao, ``{SWENet}: a
  physics-informed deep neural network ({PINN}) for shear wave elastography,''
  \emph{IEEE Transactions on Medical Imaging}, 2023.

\bibitem{pinn_material1}
Y.~Chen, L.~Lu, G.~E. Karniadakis, and L.~D. Negro, ``Physics--informed neural
  networks for inverse problems in nano-optics and metamaterials,''
  \emph{Optics Express}, vol.~28, no.~8, pp. 11\,618--11\,633, Apr 2020.

\bibitem{pinn_is1}
R.~Guo, Z.~Lin, T.~Shan, X.~Song, M.~Li, F.~Yang, S.~Xu, and A.~Abubakar,
  ``Physics embedded deep neural network for solving full-wave inverse
  scattering problems,'' \emph{IEEE Transactions on Antennas and Propagation},
  vol.~70, no.~8, pp. 6148--6159, 2022.

\bibitem{pinn_is2}
Y.-D. Hu, X.-H. Wang, H.~Zhou, L.~Wang, and B.-Z. Wang, ``A more general
  electromagnetic inverse scattering method based on physics-informed neural
  network,'' \emph{IEEE Transactions on Geoscience and Remote Sensing},
  vol.~61, pp. 1--9, 2023.

\bibitem{ie1}
K.~F. Warnick and W.~C. Chew, ``Numerical simulation methods for rough surface
  scattering,'' \emph{Waves in random media}, vol.~11, no.~1, p.~R1, 2001.

\bibitem{ie2}
A.~G. Voronovich, \emph{Wave scattering from rough surfaces}.\hskip 1em plus
  0.5em minus 0.4em\relax Springer Science \& Business Media, 2013, vol.~17.

\bibitem{mom1}
R.~F. Harrington, \emph{Field Computation by Moment Methods}.\hskip 1em plus
  0.5em minus 0.4em\relax Wiley-IEEE Press, 1993.

\bibitem{mom2}
C.~Bourlier, N.~Pinel, and G.~Kubick{\'e}, \emph{Method of moments for 2D
  scattering problems: basic concepts and applications}.\hskip 1em plus 0.5em
  minus 0.4em\relax John Wiley \& Sons, 2013.

\bibitem{kingma2014adam}
D.~P. Kingma and J.~Ba, ``Adam: A method for stochastic optimization,''
  \emph{arXiv preprint arXiv:1412.6980}, 2014.

\bibitem{pytorch}
A.~Paszke, S.~Gross, F.~Massa, A.~Lerer, J.~Bradbury, G.~Chanan, T.~Killeen,
  Z.~Lin, N.~Gimelshein, L.~Antiga, A.~Desmaison, A.~K\"{o}pf, E.~Yang,
  Z.~DeVito, M.~Raison, A.~Tejani, S.~Chilamkurthy, B.~Steiner, L.~Fang,
  J.~Bai, and S.~Chintala, \emph{PyTorch: an imperative style, high-performance
  deep learning library}.\hskip 1em plus 0.5em minus 0.4em\relax Red Hook, NY,
  USA: Curran Associates Inc., 2019.

\bibitem{automatic}
A.~Paszke, S.~Gross, S.~Chintala, G.~Chanan, E.~Yang, Z.~DeVito, Z.~Lin,
  A.~Desmaison, L.~Antiga, and A.~Lerer, ``Automatic differentiation in
  {PyTorch},'' in \emph{NeurIPS Autodiff Workshop}, 2017.

\bibitem{supporting-material}
Y.~Chen, ``Supporting material,'' URL
  \url{https://github.com/yc397/pinn_rough_surface}, 2024.

\end{thebibliography}

\end{document}